# Real Differences between OT and CRDT for Co-Editors

Chengzheng Sun, Nanyang Technological University, Singapore
David Sun, Codox Inc., United States
Agustina, Nanyang Technological University, Singapore
Weiwei Cai, Nanyang Technological University, Singapore

OT (Operational Transformation) was invented for supporting real-time co-editors in the late 1980s and has evolved to become a core technique used in today's working co-editors and adopted in major industrial products. CRDT (Commutative Replicated Data Type) for co-editors was first proposed around 2006, under the name of WOOT (WithOut Operational Transformation). Follow-up CRDT variations are commonly labeled as "*post*-OT" techniques capable of making concurrent operations *natively* commutative, and have made broad claims of superiority over OT solutions, in terms of correctness, time and space complexity, simplicity, etc. Over one decade later, however, CRDT solutions are rarely found in working co-editors, while OT solutions remain the choice for building the vast majority of co-editors. Contradictions between this reality and CRDT's purported advantages have been the source of much debate and confusion in co-editing research and developer communities. What is CRDT *really* to co-editing? What are the *real* differences between OT and CRDT for co-editors? What are the key factors that may have affected the adoption of and choice between OT and CRDT for co-editors in the real world? In this paper, we report our discoveries, in relation to these questions and beyond, from a comprehensive review and comparison study on representative OT and CRDT solutions and working co-editors based on them. Moreover, this work reveals facts and presents evidences that refute CRDT claimed advantages over OT. We hope the results reported in this paper will help clear up common myths, misconceptions, and confusions surrounding alternative co-editing techniques, and accelerate progress in co-editing technology for real world applications.

## KEYWORDS

Operational Transformation (OT), Commutative Replicated Data Type (CRDT), concurrency control, consistency maintenance, real-time collaborative editing, distributed/Internet/cloud computing technologies and systems, Computer Supported Cooperative Work (CSCW) and social computing.

## 1 INTRODUCTION

Real-time co-editors allow multiple geographically dispersed people to edit shared documents at the same time and see each other's updates instantly [1,6,12,13,14,15,35,40,50,51,56,68,72]. One major challenge in building such systems is consistency maintenance of documents in the face of concurrent editing, under high communication latency environments like the Internet, and without imposing interaction restrictions on human users [12,50,51].

Operational Transformation (OT) was invented to address this challenge [12,50,57,68] in the late 1980s. OT introduced a framework of transformation algorithms and functions to ensure consistency in the presence of concurrent user activities. The OT framework is grounded in established distributed computing theories and concepts, principally in *concurrency* and *context* theories [22,50,62,63,77,78]. Since its inception, the scope of OT research has evolved from the initial focus on consistency maintenance to include a range of key collaboration-enabling capabilities, including *group undo* [35,41,53,54,62,63], and *workspace awareness* [1,18,56]. In the past decade, a main impetus to OT research has been to move beyond plain-text co-editing [6,12, 35,40,50,51,54,57,66,67,71], and to support real-time collaboration in rich-text co-editing in word processors [56,61,64, 76], HTML/XML Web document co-editing [9], spreadsheet co-editing [65], 3D model co-editing in digital media design tools [1,2], and file synchronization in cloud storage systems [3]. Recent years have seen OT being widely adopted in industry products as the core technique for consistency maintenance, ranging from battle-tested online collaborative rich-text document editors like Google Docs[1][10], to emerging start-up products, such as Codox Apps[2].

A variety of OT-alternatives for consistency maintenance in co-editors had also been explored in the past decades [13,15,17,38,39,68]. One notable class of techniques is CRDT[3] (Commutative Replicated Data Type) for co-editors [4,5,7,23,30,36,37,38,42,43,44,73,74,75]. The first CRDT solution for consistency maintenance in plain-text co-editing appeared around 2006 [37,38], under the name of WOOT (WithOut Operational Transformation). One motivation behind WOOT was to solve the *FT (False Tie)* puzzle in OT [49,51] (further discussed in Section 4.1.2), using a radically different approach from OT. Since then, numerous WOOT revisions (e.g. WOOTO [74], WOOTH [4]) and alternative CRDT solutions (e.g. RGA [42], Logoot [73,75], LogootSplit [5]) have appeared in literature. CRDT has often been labeled as a "*post*-OT" technique that makes concurrent operations *natively* commutative, and does the job "*without operational transformation*" [37,38], and "*without concurrency control*" [23]. CRDT solutions have made broad claims of superiority over OT solutions, in terms of *correctness,* time-space *complexity*, and *simplicity*, etc. After over a decade, however, CRDT solutions are rarely found in working co-editors or industry co-editing products, but OT solutions remain the choice for building the vast majority of co-editors. In addition, the scope of nearly all CRDT solutions for co-editing are confined to resolving issues in plain-text editing, while the scope of OT has been extended from plain-text editing to rich text word processors, and 3D digital media designs, etc.

The contradictions between these realities and CRDT's purported advantages have been the source of much debate and confusion in co-editing research and developer communities. What is CRDT *really* to co-editing? What are the *real* differences between OT and CRDT for co-editors? What are the key factors that may have affected the adoption of and choice between OT and CRDT for co-editors in the real world? We believe that a thorough examination of these questions is relevant not only to researchers exploring the frontiers of collaboration-enabling technologies and systems, but also to practitioners who are seeking viable techniques to build real world collaboration tools and applications.

To seek answers to these questions and beyond, we set out to conduct a comprehensive review and comparative study on representative OT and CRDT solutions and working co-editors based on them, which are available in publications or from publicly accessible open-source project repositories. From this exploration, we made a number of research discoveries, some of which are rather surprising. One key discovery is that CRDT is similar to OT in following a *general transformation* approach to achieving consistency in real-time co-editors. Revealing the hidden transformation nature of CRDT provides much-needed clarity on *what CRDT really is to co-editing*, which in turn brings forth critical insights into the *real* differences between OT and CRDT – these differences are what ultimately contribute to the issues of correctness, complexity, efficiency and practical applicability of OT and CRDT in building real world co-editors.

In this paper, we explain *what* CRDT *really* is to co-editing, explore *what*, *how,* and *why* OT and CRDT solutions are *really* different, and examine the consequences of their differences from both an algorithmic angle and a system perspective. We know of no existing work that has made similar attempts. We focus on OT and CRDT solutions to consistency maintenance in real-time co-editing in this paper, as it is the foundation for supporting other co-editing capabilities, like group undo, and issues related to non-real-time co-editing, which we plan to cover in future papers.

The rest of this paper is organized as follows. We introduce the basic consistency maintenance issue in co-editors and layout a general transformation-based consistency maintenance approach in Section 2. We review the basic OT and CRDT approaches to realizing the general transformation approach and outline a general transformation framework for describing both OT and CRDT solutions in Section 3. Then, we delve into the different technical challenges and examine their implications on correctness, complexity and efficiency of OT and CRDT solutions in Section 4. We discuss issues and differences in applying OT and CRDT to build working co-editors in Section 5. Finally, we summarize the main discoveries and contributions of this work in Section 6.

---

[1] https://www.google.com/docs/about/
[2] https://www.codox.io
[3] In literature, CRDT can refer to a number of different data types [44]. In this paper, we focus *exclusively* on CRDT solutions *for text co-editors*, which we abbreviate as "CRDT" in the rest of the paper, though occasionally we use "CRDT for co-editors" for emphasizing this point and avoiding misinterpretation.

## 2 BASIC IDEAS OF A GENERAL TRANSFORMATION APPROACH

Modern real-time co-editors have commonly adopted a *replicated architecture*: the editor application and shared documents are replicated at all co-editing sites. A user may directly edit the local document replica and can see the local effect immediately; local edits are promptly propagated to remote sites for real-time replay there. There are two basic ways to propagate local edits: one is to propagate the edits as *operations* [12,38,50,51,73]; the other is to propagate the edits as *states* [13]. Most real-time co-editors, including those based on OT and CRDT, have adopted the *operation* approach for propagation for communication efficiency, among others. The operation approach is assumed for all editors discussed in the rest of this paper.

The central issue shared by all co-editors is this: how an operation generated from one replica can be replayed at other replicas, in the face of concurrent operations, to achieve consistent results across all replicas. Co-editors are generally required to meet three consistency requirements [51]: the first is *causality-preservation*, i.e. operations must be executed in their causal-effect orders, as defined by the happen-before relation [22]; the second is *convergence*, i.e. replicas must be the same after executing the same collection of operations; and the third is *intention-preservation*, i.e. the effect of an operation on the local replica from which this operation was originally generated must be preserved at all remote replicas in the face of concurrency.

A general approach to achieving both convergence and intention-preservation, invented in co-editing research, is based on the notion of *transformation*, i.e. an original operation is *transformed* (one way and another) into a new version, according to the impact of concurrent operations, so that executing the new version on a remote replica can achieve the same effects as executing the original operation on its local replica [51]. This approach allows concurrent operations to be executed in *different orders* (i.e. being *commutative*) but in *non-original* forms[4]. Causality-preservation can be achieved by adopting numerous suitable distributed computing techniques [12,22,50], without involving the aforementioned transformation.

The transformation approach can be illustrated by using a real-time plain text co-editing scenario in Fig. 1-(a). The initial document state *"abe"* is replicated at two sites. Under the transformation-based concurrency control, users may freely edit replicated document states to generate operations. Two operations, $O_1 = D(1)$ (to delete the character at position 1) and $O_2 = I(2,"c")$ (to insert character $c$ at position 2), are generated by User A and User B, respectively. These two operations are concurrent with each other as they are generated without the knowledge of each other [22,50]. The two operations are executed *as-is* immediately at local sites to produce *"ae"* and "*abce*", respectively; and then propagated to remote sites for replay.

In the absence of any concurrency control, the two operations would be executed in their original forms and in different orders, due to network communication latency, at the two sites, which would result in inconsistent states "*aec*" (under the *shadowed cross* at User A) and "*ace*" (at User B), as shown in Fig. 1-(a). Under the transformation-based concurrency control, however, a co-editor may execute a remote operation in a transformed form that takes into account the impact of concurrent operations, or *concurrency-impact* in short. In this example:

- At User A, $O_1$ has *left-shifting* concurrency-impact on $O_2$. So, the transformation scheme creates a new $O_2' = I(1,"c")$ from the original $O_2 = I(2, "c")$, to insert *"c"* at position 1.
- At User B, $O_2$ has no shifting concurrency-impact on $O_1$. So, the original $O_1 = O_1' = D(1)$ can be applied to delete *"b"* at position 1.

---

[4] In contrast, an alternative approach, called *serialization*, forces all operations to be executed in *the same order* and in *original forms* [12,15,50]. It has been shown the serialization approach is unable to achieve intention-preservation [51].

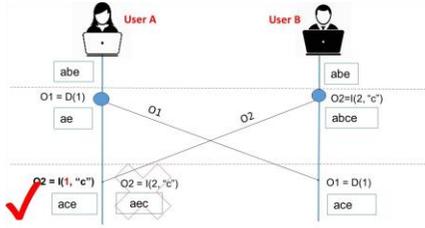
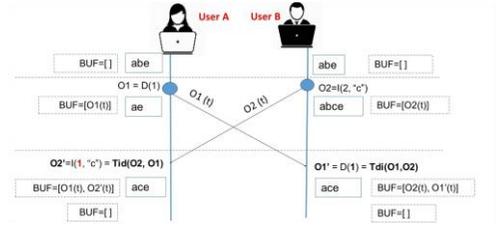

(a) Basic ideas of the general transformation.  (b) OT approach to realizing the general transformation.

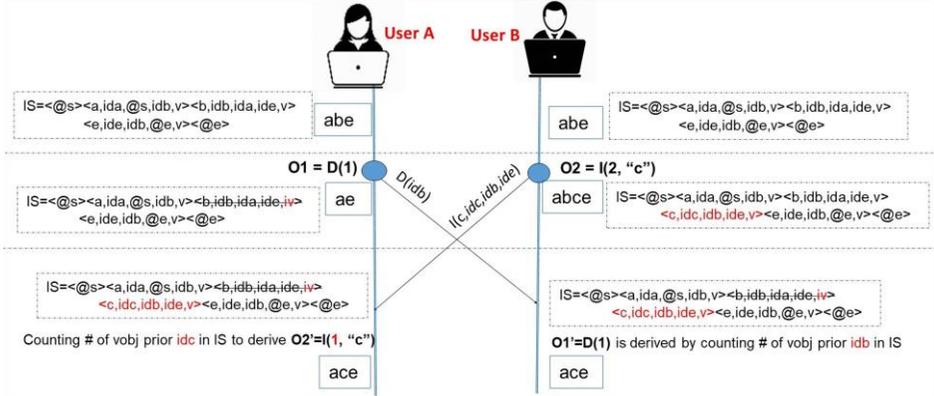

(c) WOOT (CRDT) approach to realizing the general transformation. CRDT propagates identifier-based operations.

Fig. 1. Illustrating OT and CRDT different approaches to realizing the same general transformation.

Executing $O_2'$ at User A and $O_1'$ at User B, respectively, would result in the same document state "ace", which is not only *convergent*, but also preserves the original effects of both $O_1$ and $O_2$, thus meeting the *intention-preservation* requirement [50,51]. We draw attention to the fact that, as seen in Fig. 1-(a), $O_1$ and $O_2$ are executed in *different* orders at two sites but achieve the same result, which confirms that concurrent operations are made *commutative* under the transformation-based concurrency control.

The consistency maintenance problem and solution illustrated in Fig. 1-(a) should look familiar to readers with some background in OT. Indeed it has often been used to explain basic OT ideas for consistency maintenance [12,50,51,66,67]. What might be surprising to readers is that *the same formulation of problem and solution apply equally to CRDT as well*: CRDT needs to address the same concurrency control issues, and follows the *same* general transformation approach to achieving consistency in co-editors.

Revealing the basic ideas of the general transformation approach (with more elaborations in Section 3.3) shared by both OT and CRDT at the start of our discussion helps set a common ground for examining real differences between OT and CRDT: their radically different approaches to realizing the general transformation, and the consequential correctness, complexity, and efficiency issues associated with each approach, as elaborated in the rest of this paper.

## 3 DIFFERENT APPROACHES TO REALIZING TRANSFORMATION

In the next two subsections, we introduce the basic elements of OT and CRDT, and use the same co-editing scenario in Fig. 1-(a) to illustrate and discuss how OT and CRDT realizes the general transformation. Rather than a purely algorithmic discussion, we take a system-oriented and end-to-end perspective, i.e. from the point when an operation is generated from a local editor by a user, all the way to the point when this operation is replayed in a remote editor seen by another user. We give step by step illustrations of the general process of handling an operation at both local and remote sites under both approaches, so that the subtle but key differences between OT and CRDT can be contrasted (*the devil is in the details*). In the end, we draw some key insights from the illustrations, describe the workflows of OT and CRDT under a general transformation framework, and discuss the hidden transformation nature of CRDT.

### 3.1 The OT Approach

#### 3.1.1 Key Ideas and Components

An OT solution for consistency maintenance typically consists of two key components[5]: generic *control algorithms* for managing the transformation process; and application-specific *transformation functions* for performing the actual transformation (update) on concrete operations. At each collaborating site, OT control algorithms maintain an operation *buffer* for saving operations that have been executed and may be *concurrent* with future operations.

The life cycle of a user-generated operation in an OT-based co-editor can be sketched as follows. First, an operation is generated by a user at a collaborating site. This operation is immediately executed on the local document state visible to the user. Then, this operation is timestamped to capture its concurrency relationship with other operations and saved in the local operation buffer. Next, the timestamped operation is propagated to remote sites via a communication network. When an operation arrives at a remote site, it is accepted according to the causality-based condition [12,22,50]. Then, *control algorithms* are invoked to select suitable concurrent operations from the buffer, and *transformation functions* are invoked to transform the remote operation against those concurrent operations to produce a transformed operation (a version of the remote operation is also saved in the buffer). Finally, the transformed operation is replayed on the document visible to the remote user.

For a plain-text co-editor with a pair of *insert* and *delete* operations, a total of four transformation functions, denoted as $T_{ii}$, $T_{id}$, $T_{di}$, and $T_{dd}$, are needed for four different operation type combinations [50,57,66,67]. Each function takes two operations, compares their positional relations (e.g. *left*, *right*, or *overlapping*) to derive their concurrency impacts on each other, and adjusts the parameters of the affected operation accordingly. When extending an OT solution to editors with different data and operation models, transformation functions need to be re-defined, but generic control algorithms need no change.

#### 3.1.2 A Working Example for OT

In Fig. 1-(b), we illustrate how the key components of an OT solution work together to achieve the consistent result in Fig. 1-(a). Each co-editing site is initialized with the same *external* document state *"abe"*, and an empty *internal* buffer BUF. .

*Local Operations Handling*. User A interacts with the external state to generate $O_1 = D(1)$, which results in a new state "ae". Internally, the OT solution at User A would do the following:

1. Timestamp $O_1$ to produce an *internal* operation $O_1(t)$.
2. Save $O_1(t)$ in BUF = $[O_1(t)]$.
3. Propagate $O_1(t)$ to the remote site.

Concurrently, User B interacts with the external state to generate $O_2 = I(2,"c")$, which results in a new state "*abce*". Internally, the OT solution at User B would do the following:

1. Timestamp $O_2$ to produce an *internal* operation $O_2(t)$.
2. Save $O_2(t)$ in BUF = $[O_2(t)]$.
3. Propagate $O_2(t)$ to the remote site.

---

[5] In this paper, we focus *exclusively* on OT solutions that separate *generic control algorithms* from *application-specific transformation functions* [1-3,6,9,10,12,14,24,29,32,34,35,40,41,45-54,56,61-68,70-72,76-78], as they represent the *majority* and *mainstream* OT solutions, on which existing OT-based co-editors are built. In co-editing literature, however, there are other different OT solutions (e.g. [25-28]), in which control procedures are not generic but dependent on specific types of operation and data (e.g. *insert* and *delete* operations performed on a sequence of characters), and transformation procedures may examine not only the parameters of input operations, but also concurrency relationships among other operations in the history buffer as well. Those OT solutions also adopted different criteria for consistency and algorithmic correctness, and "*control procedure and transformation functions are not separated as in previous works – instead, they work synergistically in ensuring correctness*" [28]. For details about those OT solutions, readers are referred to [25-28,57].

*Communication and Operation Propagation*: The basic OT approach described here is independent of specific communication structures or protocols (more elaboration on this point later in this paper). What is noteworthy here is that under the OT approach, operations propagated among co-editing sites are *position-based* operations.

*Remote Operation Handling*. When $O_2(t)$ arrives at User A, OT would do the following:
1. Accept $O_2(t)$ for processing under certain conditions (e.g. causal ordering [22]).
2. Transform $O_2(t)$ into $O_2'(t)$ by:
    a. invoking the control algorithm to get $O_1(t)$ from BUF, which is concurrent and defined on the same initial document state with $O_2(t)$; and
    b. invoking the transformation function $Tid(O_2, O_1)$ to produce a transformed operation $O_2' = I(1, "c")$. The *Tid* function works by comparing the position parameters 2 and 1 in $O_2$ and $O_1$, respectively, and derives that $O_2$ is performed on the right of $O_1$ in the linear document state, and hence adjusts $O_2$ position from 2 to 1 to compensate the left shifting effect of $O_1$.
3. Save $O_2'(t)$ in BUF = [$O_1(t)$, $O_2'(t)$].
4. Apply $O_2' = I(1, "c")$ on "ae" to produce "ace".

When $O_1(t)$ arrives at User B, OT would do the following:
1. Accept $O_1(t)$ for processing under certain conditions (e.g. causal ordering [22]).
2. Transform $O_1(t)$ into $O_1'(t)$ by:
    a. invoking the control algorithm to get $O_2(t)$ from BUF, which is concurrent and defined on the same initial document state with $O_1(t)$; and
    b. invoking the transformation function $Tdi(O_1, O_2)$ to produce a new operation $O_1' = D(1)$, which happens to be the same as the original $O_1$ because the *Tdi* function derives (based on the position relationship $1 < 2$) that $O_1$ is performed on the left of $O_2$ in the linear state, hence its position is not affected by $O_2$.
3. Save $O_1'(t)$ in BUF = [$O_2(t)$, $O_1'(t)$].
4. Apply $O_1' = D(1)$ on "abce" to delete "b"; the document state becomes: "ace".

There is no need to store operations in the buffer indefinitely. As soon as *there is no future operation that could possibly be concurrent with the operations in the buffer* (a general garbage collection condition for OT) [51,63,78], those operations can be garbage collected and the buffer can be reset, i.e., BUF=[].

### 3.2 The CRDT Approach
#### 3.2.1 Key Ideas and Components
WOOT [37,38] is the first CRDT solution [44] for consistency maintenance in co-editors. WOOT has two distinctive components. The first is a sequence of data objects, each of which is assigned with an *immutable* identifier and associated with either an existing character in the external document (visible to the user) or a deleted character (this internal object is then called a *tombstone*[6]). The second is the *identifier*-based operations, which are defined and applicable on the internal object sequence only.

Notwithstanding the existence of a variety of CRDT solutions, the life cycle of a user-generated operation in all CRDT solutions is the same, and can be generally sketched as follows. When a local operation is generated by a user, it is immediately executed on the document visible to the user; then this operation is given as the input to the underlying CRDT solution. The CRDT solution converts the external *position-based* input operation into an internal *identifier*-based operation, applies the internal operation in the internal object sequence, and propagates the *identifier*-based operation, to remote sites via a suitable external communication service. When a remote *identifier*-based operation is received, the CRDT solution accepts it according to certain execution conditions [22,38], applies the accepted operation to the internal object sequence, and converts the identifier-based remote operation to a *position*-based operation, which is finally replayed on the external document state visible to the user at a remote site. It is worth pointing out that the general CRDT process of handling a user-generated operation (until replaying it at a remote site) naturally existed but was often obscured in descriptions of CRDT solutions.

---
[6] To our knowledge, the AST (Address Space Transformation) solution in [17] was the first to use *marker* (tombstone-like) objects to record deleted characters in co-editors.

For WOOT to work, an insert operation carries not only the identifier of the target object (i.e. the new character to be inserted), but also identifiers of two neighboring objects corresponding to characters that are visible to the user at the time when the insert was generated. The target identifier and neighboring object identifiers, together with tombstones in the object sequence, are crucial elements in WOOT's solution to concurrency issues related to the FT puzzle [37,38].

It should be pointed out that WOOT did not (and no other CRDT solution did) actually change the formats of the external document state or operations, which are determined by the editing application (more discussion on this point in Sections 3.3 and 3.4) [8]. For consistency maintenance purpose, WOOT (and other CRDT solutions) created an additional object sequence as an internal state, identifier-based operations as internal operations, and special schemes that *convert* between external and internal operations, *search* target objects or locations, and *apply* identifier-based operations in the internal state (see more discussions on the nature of CRDT internal object sequences and operations in Sections 3.3 and 3.4).

*3.2.2 A Working Example for CRDT*

In Fig. 1-(c), we illustrate how the key functional components of WOOT work together to achieve the consistent result in a simple scenario in Fig. 1-(a). This example also serves as an illustration of the general CRDT process sketched above.

At the start, each co-editing site is initialized with the same document state *"abe"* (visible to the user), and the same internal state (IS) consisting of a sequence of objects corresponding to the initial external document state:

$$IS=<@s><a,ida,@s,idb,v><b,idb,ida,ide,v><e,ide,idb,@e,v><@e>,$$

where $<@s>$ and $<@e>$ are two special objects marking the *start* and *end* points of an internal state; each of other objects has five attributes, e.g. $<b, idb, ida, ide, v>$, where *b* is the character represented by this object, *idb* is the identifier for this object, *ida* and *ide* are the identifiers for the two *neighboring* objects, respectively, and *v* indicates the character in this object is *visible* to the user (note: *iv* indicates the character is *invisible*). An object identifier is made of two integers (*sid, seq*), where *sid* is the identifier of the site that creates the object, *seq* is the sequence number of the operations generated at that site.

*Local Operation Handling.* User A interacts with the external document to generate a position-based operation $O_1 = D(1)$, which results in a new state "ae". WOOT handles $O_1$ as follows:
 1. Convert the position-based $D(1)$ into the identifier-based $D(idb)$ by:
    a. searching the object sequence, with the index position 1 in $O_1$, to locate the target object $<b,idb,ida,ide,v>$ by counting only visible objects ($v$ = true);
    b. creating an identifier-based $D(idb)$, where *idb* is taken from $<b, idb, ida, ide, v>$.
 2. Apply $D(idb)$ to the object sequence by setting *iv* in the target object, which becomes a *tombstone* (also depicted by a line crossing the object in Fig. 1-(c)).
 3. Propagate $D(idb)$, rather than $D(1)$, to User B.

Concurrently, User B interacts with the external document to generate a position-based operation $O_2 = I(2, "c")$, which results in a new state "*abce*". WOOT handles $O_2$ as follows:
 1. Convert the position-based $I(2, "c")$ into the identifier-based $I(c,idc,idb,ide)$ by:
    a. searching the object sequence, with the index position 2 in $O_2$, to find the two visible neighboring objects between the insert position in the object sequence by counting visible objects;
    b. creating an identifier-based operation $I(c, idc, idb, ide)$, where *c* is the character to be inserted, *idc* is a new identifier for *c*, *idb* and *ide* are the identifiers of the two *neighboring* objects, respectively.

2. Apply *I*(*c, idc, idb, ide*) into the object sequence by creating a new object <*c,idc,idb,ide,v*> and injecting it at a *proper* location between the neighboring objects.
3. Propagate *I*(*c,idc,idb,ide*), rather than *I*(*2, "c"*), to User A.

*Communication and Operation Propagation:* The basic CRDT approach is independent of specific communication structures or protocols (more elaboration on this point later in this paper). What is noteworthy here is that operations propagated under the CRDT approach are *identifier-based* operations, which is different from the OT approach.

*Remote Operation Handling.* At User B, the remote operation *D*(*idb*) is handled as follows:
1. Accept *D*(*idb*) for processing under certain conditions [38,22].
2. Apply *D*(*idb*) in the object sequence by:
   a. searching the object sequence, with the identifier *idb* in *D*(*idb*), to find the target <*b,idb,ida,ide,v*> with a matching identifier; and
   b. setting *iv* to the target object (to mark it as a tombstone).
3. Convert the identifier-based *D*(*idb*) into the position-based *D*(*1*), where the position parameter 1 is derived by counting the number of (visible) objects from the target object ~~<*b,idb,ida,ide,iv*>~~ to the start of the object sequence.
4. Apply *D*(*1*) on the external state to delete "b".

At User A, the remote operation *I*(*c, idc, idb, ide*) is handled as follows:
1. Accept *I*(*c,idc,idb,ide*) for processing under certain conditions [38,22].
2. Apply *I*(*c, idc, idb, ide*) in the object sequence by:
   a. searching the sequence, with identifiers *idb* and *ide* in *I*(*c, idc, idb, ide*), to find the two *neighboring* objects; and
   b. creating a new object <*c, idc, idb, ide,v*> and injecting it at a *proper* location between the two neighboring objects.
3. Convert the identifier-based operation *I*(*c, idc, idb, ide*) into a position-based operation *I*(*1, "c"*), where the position 1 is derived by counting the number of visible objects from the new object <*c, idc, idb, ide,v*> to the start of the object sequence.
4. Apply *I*(*1, "c"*) on the external state.

Finally, both sites reach the same final external and internal states. In WOOT and its variations (WOOTO [74] and WOOTH [4]), there exists no scheme to safely remove those tombstones. In some other tombstone-based CRDT solutions (e.g. RGA [42]), a garbage collection scheme was proposed to remove tombstones under certain conditions.

### *3.3 Describing and Examining OT and CRDT under a General Transformation Framework*

The concrete co-editing scenario (in Fig. 1-(b) and (c)) is an instantiation of the *workflow* of OT and CRDT solutions under a *general transformation* approach for text co-editors, which is distilled from existing concurrency control solutions to co-editing in this work. We decompose the key steps in this workflow, from *local operation generation, handling, and propagation,* to *remote operation handling and replay*, and summarize the key actions in each step in Table 1.

#### *3.3.1 Key Components*
The general transformation framework includes the following key components:
1. The *external state and operation* models, which provide the working context of a transformation-based concurrency control solution:
   a. *ES (External State)*, which represents the text sequence for text editors;
   b. *EO (External Operation)*, which updates the text sequence, including *insert*(*p, c*) and *delete*(*p*), where *p* is a positional reference to the text sequence of ES, *c* is a character in ES (this parameter could be extended to a string of characters).
2. The external *Communication and Propagation (CP)* service, which is responsible for broadcasting operations among co-editing sites.
3. The core functional components of a transformation-based concurrency control solution:

a. *LOH (Local Operation Handler)*, which encapsulates the data structures and algorithms for handling local operations.
b. *ROH (Remote Operation Handler)*, which encapsulates the data structures and algorithms for handling remote operations.

Table 1 Describing OT and CRDT under the same general transformation framework. The shadowed blocks indicate common components shared by transformation-based concurrency control solutions for text editing.

| The General Transformation (GT) Approach | | |
|---|---|---|
| **Common external data and operation models, and general consistency requirements** | | |
| **ES (External State)** is a sequence of characters: ES = $c_0,c_1,c_2, ..., c_n$. <br> **EO (External Operation)** is a position-based operation: EO = *insert*($p, c$) or *delete*($p$). <br> **General consistency requirements:** causality-preservation, convergence and intention-preservation. <br> **General Transformation: GT(EO$_{in}$)→EO$_{out}$**, where **EO$_{in}$** is a user-generated input operation from a local document **ES$_{local}$**, and **EO$_{out}$** is the output operation to be executed on a remote document **ES$_{remote}$**. | | |
| **Work Flow** | **OT** | **CRDT** |
| **Local User** | User A interacts with the local editor to generate a *position-based* **EO$_{in}$**, which takes effects on the **ES$_{local}$** immediately and is given to the underlying LOH. | |
| **LOH** | **LOH(EO$_{in}$) → O$_t$:** <br> 1. **Timestamp** a position-based **EO$_{in}$** to make O$_t$. <br> 2. **Save** O$_t$ in the operation buffer. <br> 3. **Propagate** O$_t$ to remote sites. | **LOH(EO$_{in}$) → O$_{id}$:** <br> 1. **Convert** a position-based **EO$_{in}$** into an identifier-based O$_{id}$. <br> 2. **Apply** local O$_{id}$ in the object sequence. <br> 3. **Propogate** O$_{id}$ to remote sites. |
| **CP** | O$_t$ is *position-based*. | O$_{id}$ is *identifier-based*. |
| | A causally-ordered operation propagation and broadcasting service | |
| **ROH** | **ROH(O$_t$) → EO$_{out}$:** <br> 1. **Accept** a remote O$_t$ under certain conditions, e.g. *causally-ready*. <br> 2. **Transform** O$_t$ against concurrent operations in the buffer to produce O$_t$' and **EO$_{out}$** (= O$_t$' without timestamp). <br> 3. **Save** O$_t$ (or O$_t$') in the buffer. <br> 4. **Apply** EO$_{out}$ on ES$_{remote}$. | **ROH(O$_{id}$) → EO$_{out}$:** <br> 1. **Accept** a remote O$_{id}$ under certain conditions, e.g. *causally-ready*. <br> 2. **Apply** remote O$_{id}$ in the object sequence. <br> 3. **Convert** identified-based O$_{id}$ to position-based **EO$_{out}$**. <br> 4. Apply EO$_{out}$ on ES$_{remote}$. |
| **Remote User** | User B observes the effect of the remote **EO$_{out}$** on **ES$_{remote}$**, which has the same effect of **EO$_{in}$** on **ES$_{local}$** observed by **User A**. | |

As described in Table 1, both OT and CRDT take the same *position-based* input operation EO$_{in}$ (defined on ES$_{local}$) at the local site, and produce a transformed *position-based* output operation EO$_{out}$ (defined on ES$_{remote}$) at a remote site. The modellings of the EO as *position-based* operations and the ES as *a sequence of characters* have been commonly adopted in OT and CRDT solutions. This data and operation modelling is neither accidental nor merely a modelling convenience, but is consistent and well-supported by decades of practice of building text editors [8,11,21,69]. We should further highlight that the use of position-based operations *does not* imply the text sequence must be implemented as an array of characters. The positional reference to the text sequence could be (and has commonly been) implemented in numerous data structures, such as an array of characters, the linked-list structures, the buffer-gap structures, and piece tables, etc. [8,69].

Also shown in Table 1, OT and CRDT solutions share the same general requirement for the CP component: an external *causally-ordered* operation propagation and broadcasting service, which may or may not involve a central server (more discussions on this point later in Section 5.5). One characteristic difference in the CP component is: the propagated operations are *position-based* in

OT solutions, but *identifier-based* in CRDT solutions. In the core components LOH and ROH, however, OT and CRDT differ *significantly*, which are elaborated in the next section.

*3.3.2 Direct and Indirect Transformations*

Fundamentally, every transformation-based concurrency control solution must have a way to record the *concurrency-impact* information arising from concurrent user actions. In OT solutions, such information is recorded as *a buffer of concurrent operations*; external position-based operations are timestamped but the positional reference nature is preserved in internal operations. In CRDT solutions, concurrency-impact information is recorded in *an internal object sequence*, which maintains the effects of *all* (sequential or concurrent) operations, and external position-based operations are converted into identifier-based internal operations. These differences are described in the LOH component in Table 1.

In the ROH component, OT and CRDT use radically different methods to derive the transformed operation at a remote site. In OT solutions, when a remote position-based operation arrives, control algorithms process it against selected concurrent operations in the buffer one-by-one, and invoke transformation functions to do the transformation in each step. The actual transformation is based on a *compare-calculate* method, which *compares* numerical positions (using relations <, =, or >) between the two input operations, and *calculates* their positional differences (using arithmetic primitives + or −) to derive the position of an output operation, as illustrated in Fig. 1-(b). In CRDT solutions (e.g. WOOT), when an identifier-based operation arrives at a remote site, it is first applied in the internal object sequence, then the *transformed* (position-based) operation is derived by using a *search-count* method, which *searches* objects in the sequence and *counts* the number of *visible* objects along the way, as illustrated for WOOT in Fig. 1-(c). Logoot-like solutions have no tombstones, so all objects correspond to *visible* characters in the external state and the search-count method can be realized using binary-search which is not the same as in WOOT, but with its own special issues (as specified in [73,75]).

In summary, OT records the concurrency-impact information in a buffer of concurrent operations, and transforms position-based operations *directly* by selecting concurrent operations from the buffer, comparing and calculating positional differences between concurrent operations. In contrast, CRDT solutions record the effects of all (sequential and concurrent) operations in an internal object sequence, and transforms operations *indirectly* by converting a position-based operation into an identifier-based operation at a local site, and converting an identifier-based operation back to a position-based operation, based on searching and counting visible objects in the internal object sequence, at a remote site.

*3.4 Discussions on the Transformation Nature of CRDT*

*3.4.1 The Hidden Transformation Nature of CRDT*

When we put OT and WOOT solutions to the same co-editing example side-by-side in Fig. 1, it is clear that both solutions produce identical *position-based* operations: $O_2 = I(2, "c")$ is transformed and become $O_2' = I(1, "c")$ at User A, while $O_1 = D(1)$ is unchanged at User B. The reader can verify this by comparing Fig. 1-(c) (for WOOT) and Fig. 1-(b) (for OT). This is an example that shows WOOT indeed is an alternative to realizing the general transformation. Moreover, when we describe the CRDT approach under the general transformation framework in Table 1, the transformation nature of CRDT becomes clear as well.

Why was the transformation nature of CRDT not evident previously? We draw attention to Steps 3 and 4 in handling a remote operation in Fig. 1-(c) (and the same steps in the ROH component for CRDT in Table 1). These two steps play the roles in converting an identifier-based operation into a position-based operation, and applying a position-based operation on the external state to ensure consistency. However, both steps are omitted in the description of WOOT [38] and WOOT variations: the final step of handling a remote operation ends at Step 2, after integrating the identifier-based operation into the internal object sequence. For the scenario in Fig. 1-(c), if Steps 3 and 4 were omitted, User B would still see the document as *"abce"* even after the remote operation $D(id_b)$ has been integrated into the internal sequence, while User A would continue to see the document as *"ae"* after $I(c, id_c, id_b, id_e)$ has been internally processed. In each case, the external documents visible by Users A and B are neither *convergent* nor *intention preserving*. It is clear these steps are not mere implementation details, but crucial steps to ensure the correctness of a consistency maintenance solution for co-editing.

In WOOT [38], a *value(S)* function was briefly mentioned and supposed to map the internal object sequence *S* to the external state visible to the user. However, there was no hint on *when* and *how* the *value(S)* function might be invoked to map the internal object sequence *S* to the external document state, to accomplish the final effect of replaying a remote operation. To achieve *real-time* update of the external document, *value(S)* should be invoked whenever a remote identifier-based operation is integrated into the internal object sequence. In principle, the *value(S)* function could be implemented in two alternative ways. One is to derive a position-based operation and apply this operation to the external document, as illustrated in Fig. 1-(c). The other is to: (1) *scan* the internal object sequence to extract visible characters and generate a new sequence by character-wise concatenation, and (2) *reset* the external document state with the generated sequence of characters, which will have included the effect of the newly integrated remote operation. The second alternative is generally more expensive than the first one. One way or another, handling a remote operation must include the steps that change the external document visible to the user. We will come back later to discuss how these steps indeed manifested themselves in the documentation and/or implementation of CRDT-based co-editors built by practitioners (see Section 5.3).

*3.4.2. Myths and Misconceptions Surrounding CRDT*
In this section, we address common myths and misconceptions surrounding CRDT object sequences and operations, which had obscured *what CRDT really is to co-editors*. One common misconception is that CRDT object sequences are the text editor's internal data structures, and identifier-based operations are *native* to the editor. This leads to the illusion that there is no need for position-based operations, let alone the need to convert them to/from identifier-based operations. Evidences from existing CRDT solutions suggest otherwise: for tombstone-based WOOT variations [4,37,38,74] and RGA [42], the conversion of *local* position-based operations into identifier-based operations was *explicitly* described, although the conversion of *remote* identifier-based operations to position-based operations was omitted; for non-tombstone-based CRDT solutions, such as Logoot variations [73,75], the conversion of *remote* identifier-based operations to position-based operations was *explicitly* described, though the conversion of *local* position-based operations into identifier-based operations was obscured. Clearly, designers of CRDT solutions were cognizant of the fact that CRDT identifier-based operations and object sequences were invented for concurrency control purposes, but not native to text editors.

Furthermore, there are good reasons why CRDT object sequences and identifier-based operations are poor candidates as native data structures and operations for text editors (and by deduction text co-editors). First, data structures and operation models of text editors ought to be designed for effective and efficient support for standard text editing operations and user interactions. There exists substantial well established prior art on how to create and optimize text editors that are performant (e.g. initial loading time, memory paging speeds, etc.) [8,11,21,69] – desirable properties that should be preserved in co-editors as well. However, CRDT object sequences and identifier-based operations were created for supporting CRDT-based concurrency control only, without any concern for efficient support of standard text editing functionalities. Moreover, experiences of past decades in co-editing research suggest that co-editors should be built by separating, rather than mixing, concerns about concurrency control from concerns about conventional editing functions to allow for modularity, simplicity, and efficiency of both conventional editing functions and concurrency control solutions. In OT-based co-editors [1,51,53,56,57,64,76], for example, the choice of strategies (e.g. what native data structures or operation models to use) for implementing efficient document editing is completely left to application designers, and the support for real-time collaboration is layered above and interfaced with the editing application's exposed *abstract-data-type*, which is, in the case of text editing, a sequence of characters [8, 21]. Last but not least, existing research has found various *correctness* and *efficiency* issues with CRDT object sequences, operations and manipulation schemes for serving the *intended* concurrency control purpose (see detailed discussions in Section 4.2); it is *inconceivable* to use them as the basis for supporting *unintended* functions in standard text editors.

Closely related to above misconceptions is the idea that CRDT makes concurrent operations *natively* commutative *by design*, whereas OT makes concurrent operations commutative *after the fact* [43,44]. As revealed above, CRDT *identifier-based* operations are *not* native to text editors, but *only* commutative within the CRDT *object sequence*. This CRDT commutative property alone does not make concurrent *position-based* editing operations commutative on the text sequence visible to users, which is the real *objective* of any transformation-based concurrency control solution for co-editors. In fact, CRDT solutions make concurrent *position-based* operations commutative to co-editors *after the fact* as well, albeit *indirectly* by converting between *position-based* and *identifier-based* operations. Whether the CRDT approach to realizing the general transformation and achieving the objective of consistency maintenance is a better alternative to OT is the subject examined in follow up sections.

## 4 DIVERGING TECHNICAL CHALLENGES FOR OT AND CRDT

From dissecting and examining numerous representative OT and CRDT solutions, we have drawn some key insights on their fundamental differences in realizing the general transformation. One insight is the *direct* vs *indirect* transformation approaches (discussed in Section 3.3), and another insight is the *concurrency-centric* vs *content-centric* realization approaches, taken by OT and CRDT, respectively. The OT approach is *concurrency-centric* in the sense that an OT solution treats generic concurrency issues among operations with the *first priority* at its core by means of control algorithms, and separately handles application-specific data and operation modelling issues within transformation functions. In contrast, the CRDT approach is *content-centric* in the sense that a CRDT solution is designed around manipulating special *contents*, including an object sequence, identifiers, and schemes for searching and applying identifier-based operations in the object sequence, which are directly related to application data and operation models. In effect, CRDT solutions treat application-specific data and operation issues with the *first priority*, but mix concurrency issues within object search and manipulation schemes.

In this section, we examine how different approaches taken by OT and CRDT had resulted in different technical challenges and had major impacts on the correctness, complexity and efficiency of OT and CRDT solutions.

### *4.1 Correctness, Complexity and Efficiency Issues with OT*

#### *4.1.1 Control Algorithms and the dOPT Puzzle*

Control algorithms are at the core of the OT concurrency-centric approach. Designing correct and efficient OT control algorithms used to be a major challenge [50]. Under the first control algorithm – *dOPT* (distributed OPerational Transformation) [12], two operations could be transformed with each other as long as they had a concurrency relationship, which turned out to be inadequate. This algorithmic flaw, named *the dOPT puzzle* later, was subtle and had taken a few years for several researchers to independently discover and resolve it [50].

The key to resolving this puzzle is to ensure the two input operations to a transformation function are not only *concurrent*, but also defined on *the same document state*, or *equivalent contexts* [51]. With the guarantee of the *context-equivalence* condition, a transformation function can compare parameters of input operations to derive their concurrency-impact on each other. Detecting and resolving the dOPT puzzle has led to the invention of multiple OT control algorithms capable of ensuring the *context-equivalence* condition, and the establishment of the theory of *operation context* [50,62,63], which become a cornerstone of OT correctness. For a comprehensive review of independent solutions to the dOPT puzzle, the reader is referred to [50].

Some OT control algorithms, including adOPTed [40], GOT [49,51], GOTO [50], and SOCT2 [47], had the *quadratic* time complexity in transforming a remote operation – $O(c^2)$, where $c$ is the number of concurrent operations involved in transforming an operation. Though this theoretic complexity was not a concern in real-time co-editors because typically c ≤ 10 [57], more efficient control algorithms were proposed (e.g. Jupiter [34], NICE [45], TIBOT [24,78], SOCT4 [71],

Google Wave and Docs OT [10,72,32], COT [62,63], and POT [78]), with a time complexity of $O(c)$ for transforming a remote operation. For processing local operations, OT solutions generally have the constant time complexity $O(1)$[7].

For recording concurrency-impact, an OT solution keeps a buffer of operations. The space complexity of this buffer is characterized by two variables $c$ and $m$, where $c$ is the same as above, and $m$ is the number of users in a session, and depends on whether using scalars or vectors for timestamping operations, and whether maintaining single or multiple transformation paths in the buffer, as summarized in Table 2.

Due to its *concurrency* nature, the variable $c$ has two properties: (1) $c$ is often bounded by a small value, e.g. $c \leq 10$, in real-time co-editing sessions, in which the number ($m$) of users is small (e.g. $m \leq 5$), and the number of operations each user may generate per second is small (e.g. $\leq 2$) due to the relative slow pace of human interactions and operation composing schemes commonly used in real-time co-editors [57,72]; (2) $c$ can be reduced to *zero* by applying garbage collection schemes that remove buffered operations whenever it is no longer possible for them to be concurrent with future operations [51,57,63,78]. Garbage collecting non-concurrent operations has been well-established not only in theory, but also commonly adopted in OT-based co-editors, including Google Wave and Docs [10,72], CodoxWord [56], Codox Apps, etc.

Table 2. Space complexities in representative (not exhaustive) OT solutions

|  | **Single-T-Path** | **Multi-T-Path** |
|---|---|---|
| **Scalar Timestamps** | $O(c)$[10,24,34,45,71,72] | $O(c*m)$ [78] |
| **Vector Timestamps** | $O(c*m)$[12,47,49,50,51] | $O(c*m^2)$ [40,63] |

*4.1.2 Transformation Functions and the False-Tie Puzzle*

For an OT solution to work, transformation functions need to preserve some transformation properties under certain conditions (e.g. what control algorithms are used in the OT solution).

**Convergence Property 1 (CP1):** Given $O_a$ and $O_b$ defined on the document state *DS*, and a transformation function *T*, if $O_a' = T(O_a, O_b)$, and $O_b' = T(O_b, O_a)$, the following holds:
$$DS \circ O_a \circ O_b' = DS \circ O_b \circ O_a',$$
which means applying $O_a$ and $O_b'$ in sequence on *DS* produces the same state as applying $O_b$ and $O_a'$ in sequence on *DS*.

Preserving CP1 is the key for OT to make concurrent editing operations *commutative*. Numerous transformation functions capable of preserving CP1 (under text and other document models) have been designed [1,2,3,40,65,66,67]. Further research found that CP1 alone may not be sufficient to ensure convergence (for OT solutions supporting more than 2 users); an additional property CP2 may be required under certain conditions (more elaboration on those conditions in Sections 4.1.3 and 4.3.1) [40,63,78].

**Convergence Property 2 (CP2):** Given $O_a$, $O_b$ and $O_c$ defined on the same state, and a transformation function *T*, if $O_c' = T(O_c, O_b)$ and $O_b' = T(O_b, O_c)$, the following holds:
$$T(T(O_a, O_b), O_c') = T(T(O_a, O_c), O_b'),$$
which means transforming $O_a$ against $O_b$ and then $O_c'$ produces the same operation as transforming $O_a$ against $O_c$ and then $O_b'$.

---

[7] The local operation processing time covers the period during which the local operation is timestamped and saved in the buffer (ready for propagation), as sketched in OT LOH in Table 1. For most OT solutions, a local operation in the buffer is propagated as-is without further processing. For a few OT solutions, notably Google Wave and Docs OT, TIBOT, and SOCT4, local operations may wait in the buffer until certain conditions are met, e.g., a local operation is not propagated until the prior propagated local operation has been acknowledged (for Google Wave and Docs OT); a local operation is not propagated until all remote operations with *time-interval-based* timestamps (for TIBOT) or *global sequence numbers* (for SOCT4) that are earlier than that of the waiting local operation have been received and processed. While waiting in the buffer, a local operation may be transformed with incoming remote operations, and such processing time is part of handling a remote operation by ROH (with the complexity $O(c)$). Handling a local operation is completed at the moment when an OT solution is able to handle another local operation (by LOH) *or* remote operation (by ROH).

It is worth highlighting that the transformation function *T*, document state *DS*, and operation *O* are all *unspecified* in CP1 and CP2 specifications, meaning CP1 and CP2 are applicable to transformation functions defined for any document states and operation models. Establishing general transformation conditions and properties for correctness, including CP1 and CP2, is one of major achievements in past OT research [12,35,40,50,51,57,62,63,66,67,78].

Unlike CP1, which is relatively easy to preserve, CP2 is non-trivial and could be violated under certain circumstances. The first CP2-violation case in text editing was reported in [49], late named the *False-Tie (FT)* puzzle as it always involves an *insert-insert-tie* that does not exist in original user-generated operations but occurs only between transformed operations [57]. It turned out that detecting and resolving CP2-violation and the FT puzzle had enormous impact on follow-up development of OT (with respect to CP2-correctness), the invention of WOOT, and other OT alternatives as well (see more discussions in Section 4.4.2) [19,37,38,39].

*4.1.3   OT Solutions to the CP2 Issue and the FT Puzzle*
One fundamental point we should highlight at the start is that CP2 is *not* unconditionally required, but required for transformation functions *only if* the control algorithms in the OT solution transform concurrent operations in *arbitrary* orders, or more precisely, transform the same pair of concurrent operations in different contexts [40,57,62, 63,78]. This CP2 *precondition* has played a key role in the invention and verification of OT solutions to the CP2 issue along two alternative paths.

First, the CP2 issue can be addressed by designing OT control algorithms that *avoid* transforming operations in arbitrary orders (the *CP2-avoidance* approach, which is generic and independent of operation types and data models). When control algorithms with the *CP2-avoidance* capability are used in an OT solution, CP2 is no longer a correctness requirement for corresponding transformation functions. Control algorithms capable of avoiding CP2 are numerous [10,24,32,34, 45,46,62,63,71,72,78]. It should stressed that CP2-avoidance solutions impose no restriction on users' ability to edit the shared document freely and concurrently, and were designed with a variety of communication structures and protocols, with or without using a server for operation transformation and propagation, and using either vectors or scalars for timestamping operations. Therefore, it is incorrect to label CP2-avoidance solutions as being unsuitable for deployment in *peer-2-peer* co-editing environments (e.g. [4]). These points will be further elaborated in Section 5.5. A comprehensive study of CP2-avoidance control algorithms is available in [77,78].

Second, the CP2 (and FT) issue can also be addressed by designing transformation functions capable of preserving CP2 (*without* changing underlying operation and data models). This *CP2-preserving* approach is suitable to OT solutions with control algorithms that transform operations in arbitrary orders (e.g. adOPTed [40], GOTO [50], and SOCT2 [47]). Obviously, transformation functions capable of preserving CP2 (and CP1) can also be used in combination of control algorithms with the *CP2-avoidance* capability. Transformation functions capable of preserving CP2 (and CP1) for string-wise data and operation models have been designed and verified in [67].

The CP2-avoidance approach is often favored as it is generally applicable to data and operation models for a range of different applications beyond plain-text editing. OT solutions for advanced co-editing applications, such as  2D spreadsheets [65], 3D digital media design systems [1,2], and shared workspaces in cloud storage systems [3], have all used a combination of transformation functions for preserving CP1 (*plus* application-specific *combined effects* for concurrent operations [1,3,57,66]) and control algorithms for avoiding CP2[63,77,78].

There were other attempts to solve the FT puzzle and the CP2-violation issue [19,20,25,26,27, 28,37,38,39,48], which did not follow above *CP2-avoidance* or *preserving* strategies. We will discuss them, together with misconceptions surrounding FT and CP2, in Section 4.3.

*4.1.4.   Summary of Correctness, Complexity and Efficiency of OT Solutions*
In over two-decades, numerous OT solutions have been invented to address OT-special technical challenges, including the dOPT puzzle and the FT puzzle, and to support building real world co-editors from text editing to more complex editing applications. The *correctness* of key OT components, including generic control algorithms and transformation functions for a range of commonly used operation and data models (e.g. string-wise plain-text editing and beyond) [1,3,65,

66,67], has been established under well-defined conditions and properties [12,29,35,40,50,51,54, 57,63,67,78]. State-of-the-art OT solutions have achieved the *space complexity* $O(c)$ or $O(c*m)$, and the *time complexity* $O(1)$ and $O(c)$ for processing local and remote operations, respectively.

### *4.2 Correctness, Complexity and Efficiency Issues with CRDT*

In over one decade, CRDT has evolved from the initial WOOT to a myriad of CRDT solutions for plain-text co-editing [4,5,7,23,30,36,37,38,42,43,44,73,74,75]. CRDT solutions have been commonly designed around internal object sequences, object identifiers, identifier-based operations, and schemes for searching and manipulating internal object sequences, which we characterized as the *content-centric* approach. However, concurrency issues are *inherent* and *unavoidable* in unconstrained co-editing (both OT and CRDT aim to support), so CRDT solutions have to face and handle concurrency issues in *content-specific* ways, meaning to mix concurrency solutions within specific object sequences, identifiers, and associated search and manipulation schemes, leading to CRDT-special *correctness, complexity* and *efficiency* problems.

#### *4.2.1 Big C/$C_t$ Complexities in CRDT Solutions*

A direct consequence of the CRDT content-centric approach is that time and space complexities of CRDT solutions are inherently determined by a big $C$ (for *Contents*) or $C_t$ (for *Contents with tombstones*) variable − the number of objects in the object sequence. $C$ often takes a big value, e.g. $10^3 \leq C \leq 10^6$, for common text document sizes ranging from 1KB to 1MB; $C_t$ is much bigger than $C$ due to tombstones (more elaboration below).

To achieve consistency in the presence of *concurrent* inserts between two visible neighboring objects, WOOT used a recursive *IntegrateIns* procedure to search the object sequence in nested loops in order to determine the correct target location, which had the time complexity of $O(C_t^3)$ [37,38]. This recursive procedure was quite intricate and had taken multiple iterations to get it to work, as reported in [37]. The complication here can be attributed primarily to WOOT's content-specific ways of handling concurrent operations between two neighboring objects.

$O(C_t^3)$ was obviously too costly for executing a single insert operation (at both the local and remote replicas) in WOOT, so WOOT improvements were proposed, including WOOTO[74], which used a *degree* scheme to capture the relative ordering of concurrent object creations and save one round of object sequence search; and WOOTH [4], which used a hash scheme to speed up the search of neighboring objects. Both WOOTO and WOOTH achieved the time complexity $O(C_t^2)$.

RGA (Replicated Growable Array) was another CRDT solution for co-editing [42], which also adopted the tombstone-based approach to representing the internal object sequence, but used a hash scheme and vector-based total ordering which respects causal-ordering to speed up the search of the target object or location in the object sequence, and achieved the time complexity of $O(C_t)$ (or $O(C)$ with tombstone garbage collection, discussed below) for executing an insert operation, but with additional costs in maintaining the hash table and issues related to vector clocks [37,38,73].

The cost of applying an identifier-based operation in the object sequence (e.g. the cost of *IntegrateIns*) is only one part of handling a local or remote operation in WOOT (and its variations). Another often-occurring cost is searching the object sequence, which has the time complexity $O(C_t)$, and may occur at multiple places, e.g. converting a position-based operation into an identifier-based operation at the local site (using the *position* as the search key), checking the existence of an object in the sequence for determining the acceptance of a remote identifier-based operation (using the *identifier* as the key), and converting an identifier-based operation to a position-based operation for replay at a remote site (counting the number of visible objects in the sequence), etc. A more recent paper [7] acknowledged the WOOT (and other CRDT) performance problem in the *upstream* processing (i.e. processing local operations) [8], which had adverse effect on local responsiveness to users, and attributed this problem to the time complexity $O(C_t)$ for the search process in converting a position-based operation into an identifier-based operation. The proposed solution was to add one extra binary tree to speed up the local search process. With this extra tree, together with its additional complications and costs, the complexity of this local search process could be reduced to $O(log(C_t))$, which improves the upstream execution time by *several orders of magnitude* as claimed in [7]. For comparison, OT solutions commonly have the complexity $O(1)$ for the whole process of handling a local operation.

---

[8] In [7], the authors acknowledged "*upstream execution − and thus responsiveness of CRDT algorithms often performs poorly,*" but still claimed *"downstream execution of CRDT algorithms is more efficient by a factor between 25 to 1000 compared to representative OT algorithms"* [4], which is examined in Section 4.3.

Table 3. Key features of representative (not exhaustive) CRDT solutions. $C_t$ and $C$ represent the number of objects in the object sequence with or without tombstones, respectively. Time complexity is for handling one insert internally, which incurs at both local and remote sites.

| | WOOT [37,38] | WOOTO [74] | WOOTH [4] | RGA [42] | Logoot [73] | Logoot-Undo [75] |
|---|---|---|---|---|---|---|
| Object sequence | @s, o₁, o₂, …, oₙ, @e where $o_x$ is an object corresponding to a character in external document or a tombstone; @s and @e are special objects marking the start and end of the object sequence. | | | | | |
| | **tombstones** kept in the object sequence | | | | **no tombstone** kept | |
| Identifier-based ops | I(*ids, c*) and D(*id*); *c* is a character | | | | | |
| Object | <*idc,idp,idn, v,c*> **26B** | <*idc, deg, v,c*> **14B** | <*idc,deg,nlink,hlink, v,c*> **22B** | <*idc,idp,nlink, hlink, v,c*> **42B** | <*id*> an *object* has a variable size bounded by *C* | |
| Id | <*sid, seq*> **8B** | | | <*sn,sum,sid,seq*> **16B** | <*i₁,sid*>,… <*iₙ,sid*>, seq | <*i₁,sid,seq*> … <*iₙ,sid,seq*> |
| | | | | | each *id* has a variable length bounded by *C* | |
| Ids | <*idp,idc,idn*> **24B** | <*idp, idc, idn, deg*> **28B** | | <*idc, idp*> **32B** | NA | |
| Id ordering | *ids are totally ordered* | | | | | |
| | without additional constraint | | | + *causal order* | + *positional order* | |
| Timestamp | *Scalar*-based | | | *Vector*-based | NA (but require causally ordered broadcast) | |
| Time complexity | $O(C_t^3)$ | $O(C_t^2)$ | $O(C_t^2)$ | $O(C)$ with G.C. | $O(C)$ for local operations; $O(C \circ log(C))$ for remote operations. | |
| Space complexity | $O(C_t)$ | | | $O(C)$ with G.C. | between $O(C)$ and $O(C^2)$ | |

Apart from time complexities, space complexities of CRDT solutions are also characterized by the big $C_t$ or $C$ variable. The internal object sequence has the space complexity $O(C_t)$ for CRDT solutions that maintain tombstones (e.g. WOOT variations), or between $O(C)$ and $O(C^2)$ for those without maintaining tombstones (e.g. Logoot [73]). In Table 3, we give a summary of objects, identifiers, time and space complexity, and other characteristic features of some representative CRDT solutions. The actual size of each object in the sequence deserves attention: for each character of one byte in a text document, the corresponding internal object may have a size of 14, 22, or 26 bytes for WOOT variations, 42 bytes for RGA, or a variable size (bounded by *C*) for Logoot variations. This means that the space over head of a CRDT object sequence is at least 14 to 42 times larger than the original size of the external document, without counting tombstones.

### 4.2.2 Tombstone Overhead in WOOT Variations and RGA

While WOOT variations and RGA are different from each other in concrete representations of internal object sequences and operations, they have one thing in common: the use of tombstones to represent deleted objects in the object sequence. The very need for tombstones was to support identifier-based search of the object sequence in the face of *concurrent* operations. The major problem with tombstones is overhead. Here is a telling example of the unacceptable costs caused by tombstones [73], reported directly by WOOT researchers:

> "For the most edited pages of Wikipedia, the tombstone storage overhead can represent several hundred times the document size. Tombstones are also responsible of performance degradation. Indeed, in all published approaches, the execution time of modification integration depends on the whole document size – including tombstones. ... While the 'George W. Bush' page contains only about 553 lines, the number of deletions is about 1.6 million. As a consequence, tombstones-based systems are not well-suited for such documents since we obtain 1.6 million tombstones for only 553 lines."

While the prohibitive tombstone costs ($C_t = C \times 2894$ in this example) quoted above are not necessarily generalizable, the tombstone overhead was indeed a major and inherent issue with WOOT variations and RGA. Various attempts were made trying to remove tombstones as garbage. In RGA [42], a garbage collection scheme was proposed, which was based on the following conditions: a tombstone object can be removed only if: (1) no future operation will be concurrent with the *delete* operation that converted the object into a tombstone (similar to the garbage collection condition and the vector-clocks-based scheme in GOT [51]); and (2) no future operation will have a total order that is earlier than the total order of the object immediately after the tombstone in the object sequence (an additional condition related to RGA's special way of maintaining and using tombstones). However, WOOT variations have no garbage collection schemes of their own, and also cannot adopt the garbage collection scheme from RGA due to the absence of vector-clocks and different ways of maintaining and using tombstones and neighboring objects in WOOT variations under complex *concurrent* co-editing scenarios [4,38,74].

*4.2.3  Variable and Lengthy Identifiers in Logoot*

As the tombstone issue could be a road blocker for WOOT variations to be applicable to real world co-editors, an alternative CRDT solution, named Logoot [73], was proposed to avoid using tombstones in the object sequence, but impose a special *positional ordering* constraint on identifiers: *object identifiers must have a total ordering that is consistent with the positional ordering of the corresponding objects in the object sequence*.

As shown in Table 3, all CRDT solutions, with or without tombstones in object sequences, have imposed a *total ordering* on identifiers for various purposes, e.g. to determine the location of an *insert* operation when multiple concurrent operations are inserting *at the same location* in the object sequence (for WOOT variations and RGA). The *positional ordering* constraint for Logoot (and its variations) identifiers is quite special, which is to determine the location of *all* operations (being *insert* or *delete*, *concurrent* or *sequential*) in the object sequence (at remote sites) ─ the key to Logoot's correctness and complexity, and hence worth further elaboration.

For any two objects A and B in the object sequence, with identifiers *ida* and *idb*, respectively, if A is positioned before B in the object sequence, *ida* must be ordered before *idb* in the identifier ordering as well. This property is essential for searching target objects or locations in the object sequence, without the help of tombstones, in the face of concurrent editing. For Logoot to work, the key is to devise identifiers that really possess *uniqueness*, *immutability* and *positional ordering* properties, which turned out to be a major challenge.

The basic idea in the Logoot identifier scheme is to assign each object with an integer in a number system of a chosen base (e.g. $2^{64}$) according to the object's position in the object sequence. For any two objects A and B, if A precedes B in the object sequence, the integer assigned to A must be smaller than that to B. To insert a new object X between two existing objects A and B with integers *p* and *q* in their identifiers, respectively, the identifier scheme will assign object X with an integer *i*, which is *randomly* taken between *p* and *q*, such that $p < i < q$.

One issue with this basic scheme stems from the nature of *concurrent* editing: *what if* two users are inserting between the same pair of objects and the identifier schemes at the two sites generate the same random integer *i*? To break the *concurrent-insert-tie*, Logoot coupled the random integer with a site identifier (*sid*) to make a tuple <*i, sid*>. This extension, however, was still not enough.

Another issue with the identifier scheme is related to the limited space between any two integers: *what if* a user inserts a large number of objects between two existing objects, and the required integers for identifying those objects exceed the available integers between the identifiers of neighboring objects? This led to further extensions of the Logoot identifier from a single tuple to a *variable* number of tuples, which is bounded by the big *C,* and an additional local operation sequence number (*seq*) was added to maintain *uniqueness* (see Table 3).

The variable and potentially lengthy identifiers could lead to high space and time costs in the object sequence representation and searching [73]. Average lengths of identifiers depend on editing patterns, and the worst case complexity of Logoot identifiers is $O(C)$ [73,75]; the space complexity of the object sequence is $O(C^2)$; the time complexity for handling a local insert is $O(C)$, but $O(C \circ log(C))$ for a remote insert [4,73,75].

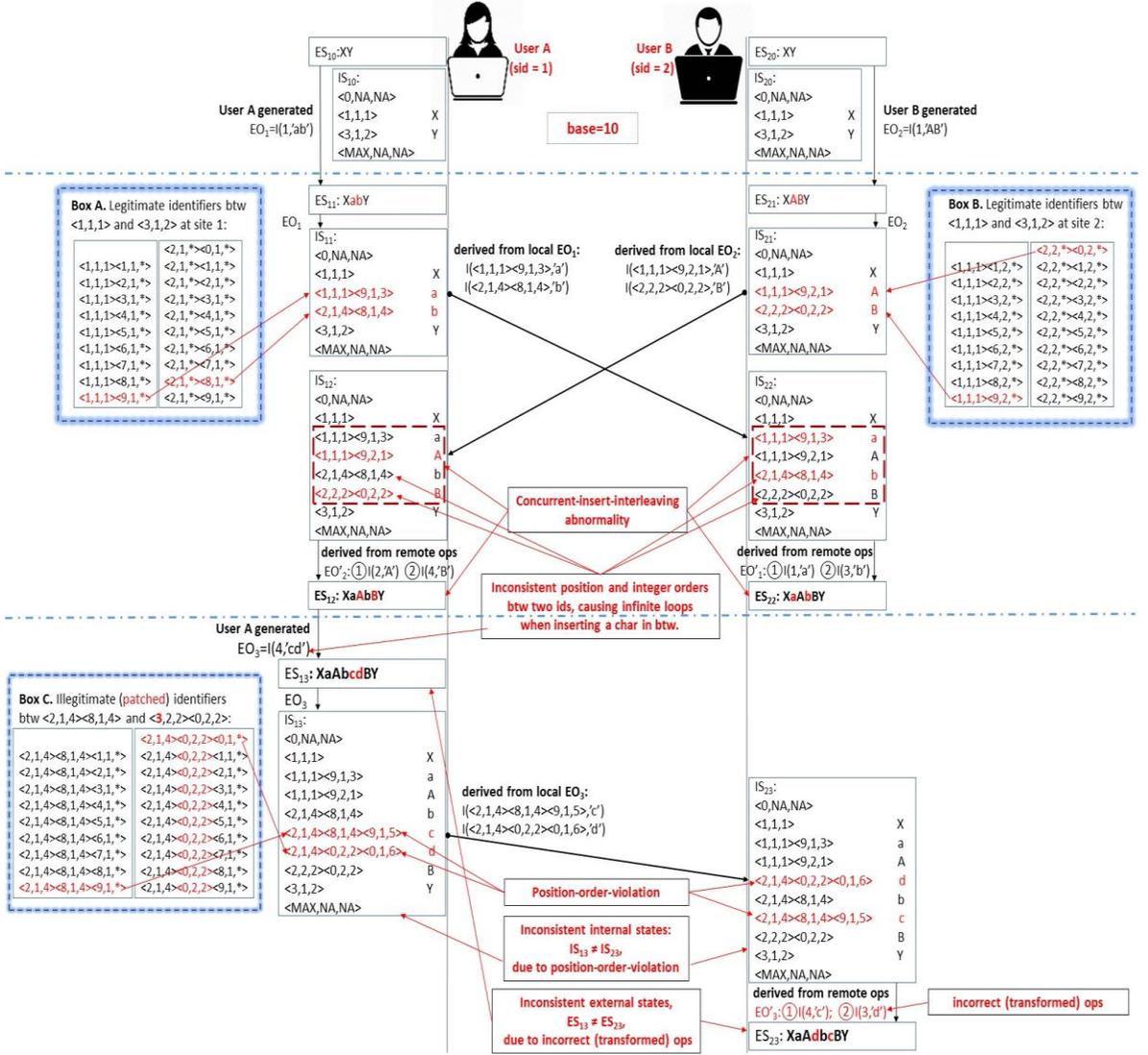

Fig 2. Illustration of multiple issues, including *concurrent-insert-interleaving*, *inconsistent-position-integer-ordering* (and *infinite loop* flaw), and *position-order-violation*, in Logoot.

To get the lengthy identifier issue under control, various patches were added to the Logoot identifier scheme, leading to increasingly complicated schemes. These patches can generate identifiers violating the positional ordering requirement, which manifests as inconsistencies in (both internal and external) document states.

In Fig 2, we use one scenario with three operations to illustrate multiple issues in the Logoot identifier scheme. For convenience, we introduce the following notations: ES for External State (visible to users); EO for (position-based) External Operation (generated by users), and IS for Internal State (the object sequence manipulated by Logoot). In the following examples, we use the identifier generation scheme described in [73,75] (both versions are basically the same). For simplicity, we use 10 as the integer base for identifiers, MAX as a symbolic value larger than the biggest integer (9) in the base, <0, NA,NA> and <MAX,NA,NA>, which were also used in [75], as two special identifiers marking the start and end of the object sequence. Initially (the portion above the first horizontal dashed line in Fig 2), the external state contains two characters "XY", which correspond to the two objects with identifiers <1,1,1> and <3,1,2>, respectively. In each tuple of an identifier, the first number is an integer, the second number is the site identifier, and the third number is the sequence number.

*4.2.4 Concurrent-Insert-Interleaving Puzzle*

Among all issues with the Logoot identifier scheme, the most critical one is the *random interleaving* anomaly, which could occur whenever two users *concurrently* insert continuous characters between the same pair of existing characters.

As shown in Fig 2, User A generates $EO_1 = I(1, "ab")$ to insert two characters "ab" between "X" and "Y"; and concurrently User B generates $EO_2 = I(1, "AB")$ to insert two characters "AB" between "X" and "Y". Internally, Logoot at the User A site first uses the position 1 in $EO_1$ to find the insert location and obtain two neighboring identifiers <1,1,1> and <3,1,2> in $IS_{10}$. Since there exists only one available integer between 1 and 3 (in the first tuples of the two neighboring identifiers), a new tuple has to be added to generate two new identifiers for the two inserted characters. According to the identifier generation scheme in [73,75], a range of legitimate identifiers (shown in **Box A** in Fig 2) are available for random choices. Without losing generality, we choose two identifiers <1,1,1><9,1,3> and <2,1,4><8,1,4> from this range to identify the two inserted characters. Then, Logoot inserts these two identifiers at the location 1 in the object sequence. Afterwards, two identifier-based operations are propagated to User B. At the User B site, a similar process occurs, two identifiers <1,1,1><9,2,1> and <2,2,2><0,2,2> are generated (from **Box B** in Fig 2) to identify the two inserted characters, and two identifier-based operations are propagated to User A.

After receiving identifier-based operations from a remote site, Logoot uses the identifiers in remote operations to determine the insert locations in the object sequence. After applying the identifier-based concurrent operations at each other sites, the combined final (both internal and external) states are consistent, but the four characters and their corresponding internal identifiers are *interleaved*: the external document state becomes "XaAbBY", rather than "XabABY" (or "XABabY"), which would normally be expected by users.

It should be highlighted that the four identifiers are legitimate identifiers (according to [73,75]), and meet the *uniqueness*, *immutability* and *position-ordering* properties required by Logoot, but their orders are *randomly interleaved*. This anomaly is rooted in Logoot's fundamental random positional identifier generation scheme.

*4.2.5 Inconsistent-Position-Integer-Ordering Puzzle*

Before the second horizontal dashed line in Fig 2, the concurrent-insert-interleaving abnormality has manifested itself, but there is another hidden problem inside the internal states (both $IS_{12}$ and $IS_{22}$): the two adjacent identifiers $id_1 = <2,1,4><8,1,4>$ and $id_2 = <2,2,2><0,2,2>$ have an *inconsistent-position-integer-ordering* problem, meaning their positional order ($id_1 < id_2$) is inconsistent with their integer order (28>20). This hidden problem would manifest as an *infinite loop* in the Logoot identifier generation scheme when any user inserts a character between these two adjacent objects. For example, when User A generates $EO_3 = I(4, "cd")$ to insert two characters "cd" at position 4, i.e. between "b" and "B" in the external state $ES_{12}$, as shown in Fig 2, Logoot would fail to generate identifiers.

In general, the Logoot identifier generation scheme will run into an *infinite loop* and *fail* to generate any new identifier between two neighboring identifiers with position integers *p* (for *left* identifier) and *q* (for *right* identifier) if $p \geq q$ (or their position and integer orders are *inconsistent*), which could easily occur when two users are inserting at the same location *concurrently*.

The infinite loop flaw was in the first Logoot paper [73], remained the same in a late version [75], and never corrected in late CRDT publications. However, we found several *patches* introduced to avoid the infinite loop in open source codes[9,10] that implemented the Logoot identifier scheme. Nevertheless, none of those patches could really solve the problem without running into new problems, which are illustrated in the next subsection.

*4.2.6 Position-Order-Violation Puzzle*

We use the patch in the Logoot library (referred to in *footnote* 9) implemented by Logoot authors [7], to illustrate the *position-order-violation* puzzle, which is caused by the patches introduced to resolve the infinite loop flaw, as shown under the second horizontal dashed line in Fig 2.

When User A generates $EO_3$ = I(4, "cd") to insert two characters "cd" between "b" and "B", the Logoot library would avoid the infinite loop by changing the *right* neighbor identifier from <**2**,2,2><0,2,2> into <**3**,2,2><0,2,2> (only inside the identifier generation algorithm, not in the internal state), to allow the use of a range of *illegitimate* identifiers from <2,1,4><8,1,4><1,1,*> to <2,1,4><0,2,2><9,1,*>, shown in **Box C** in Fig 2. However, this patch could cause numerous *abnormal* cases that cannot be handled by the original identifier generation scheme (e.g. the *ConstructId* function in [75]), which in turn requires additional patches to deal with. For example, the tuple <0,2,2> in identifiers <2,1,4><0,2,2><*,1,*> is inherited from the corresponding tuple in the *right* neighboring identifier <**3**,2,2><0,2,2>, and this inheritance is forced by one patch in the Logoot library[11]. Unfortunately, the *position-order-violation* problem manifests itself among these *illegitimate* identifiers, e.g. between the two identifiers <2,1,4><8,1,4><9,1,5> and <2,1,4><0,2,2><4,1,6> (and among other pairs as well) in this range. Trouble would occur when these two identifiers are assigned to the two new objects for the two characters "cd".

This position-order-violation does not immediately cause a trouble at the local site as the local insertion position is determined by the position number 4, rather than by these identifiers; the trouble occurs when Logoot at the remote User B site uses these identifiers to determine their positions in $IS_{22}$, and inserts "c" and "d" at corresponding positions, which results in inconsistent internal states, i.e., $IS_{13} \neq IS_{23}$, which in turn leads to *incorrect* external (transformed) operation $EO_3$', which finally results in *inconsistent* external states, i.e. $ES_{13} \neq ES_{23}$, as shown in Fig 2.

It should be pointed out that the *inconsistent-position-integer-ordering* problem (and the associated *infinite loop* flaw), and the *position-order-violation* puzzle could occur under numerous circumstances, e.g. the reader can use different identifier combinations in **Boxes A, B** and **C** in Fig 2 to create varieties of similar puzzles.

It remains a critical open issue for Logoot to find and prove a correct identifier scheme. Logoot variations, such as LogootSplit [5] and LogootUndo [75], tried to extend Logoot from supporting character-wise to string-wise operations and from supporting *do* to *undo*. Unfortunately, new identification schemes for *string* operations and *undo* have even higher complexity than that for *character-wise* operations and *do*-only (see Table 3), and their correctness was not verified either.

In general, it remains an open challenge to verify the correctness of key components in CRDT solutions, e.g. object identifiers, sequences, searching and manipulation schemes, based on well-defined criteria (yet to be established as well). Consistency (e.g. convergence) claims of a solution cannot be drawn from the assumption that all components worked according to what were *specified* or *required* (e.g. Logoot identifiers are required to possess the *positional ordering* property), but must be based on what were actually *designed*, which could be flawed. For example, the designed Logoot identifier scheme actually run into infinite loops and failed to generate any identifier, or failed to preserve the positional ordering property and led to inconsistency, as illustrated in Fig 2.

*4.2.7 Summary of CRDT Correctness, Complexity and Efficiency*

One motivation of the first CRDT solution (WOOT) was to address the FT puzzle and the CP2-violation issue in OT, which have been solved under the OT approach. As an OT-alternative to concurrency control in co-editing, CRDT solutions did the job *without* using OT algorithms, but *with* CRDT-special object sequences, identifier-based operations, and schemes for manipulating such sequences and operations, which came *with* CRDT-special correctness and complexity issues (see time and space complexities of representative CRDT solutions in Table 3). Some CRDT issues (e.g. *tombstone garbage* for WOOT variations, *inconsistent-position-integer-ordering* and *position-order-violation* for Logoot variations, etc.) are open for resolution, but others (e.g. big $C/C_t$ *complexities in time and space* for all CRDT solutions, and *concurrent-insert-interleaving* for Logoot variations, etc.) are inherent to the basic approaches taken by CRDT solutions, and it is unclear whether they can be solved without fundamental changes to these basic approaches.

---

[9] https://github.com/coast-team/replication-benchmarker.
[10] https://github.com/rudi-c/alchemy-book. It is worth noting that the author of this work also detected various issues in Logoot and pointed out Logoot "*missing what I think are key details on how to handle certain edge cases*", and devised his own patches to deal with those missing key details. Those patches also came with problems which may result in state inconsistencies or system crashes. Detailed analysis of those issues is beyond the scope of this article.
[11] This patch is implemented in functions *generateLineIdentifiers* and *constructIdentifier* in https://github.com/coast-team/replication-benchmarker/blob/master/src/main/java/jbenchmarker/logoot/.

*4.3 Facts and Myths about Correctness, Complexity and Simplicity of OT and CRDT*

In this section, we review and dispel a number of common misconceptions and controversies about the correctness, complexity and simplicity of OT and CRDT solutions.

*4.3.1. Misconceptions in Evaluating OT Correctness*

Achieving convergence is one key requirement for OT solutions [12, 50,51]; two transformation properties CP1 and CP2, among others, are directly relevant to achieving convergence. In [40], one theorem [12] established that CP1 and CP2 are two *necessary* and *sufficient* conditions to achieve *convergence* under the control of the adOPTed algorithm. This CP1-CP2-theorem is applicable to *transformation functions* in OT solutions that allow concurrent operations to be transformed in *arbitrary orders* or *under different contexts* [63]. This theorem is important, but unfortunately often misinterpreted, which is a source of various misconceptions surrounding OT correctness, particularly CP2 correctness, which is collectively called *the CP2 syndrome*.

The top symptom of this syndrome is to misinterpret CP1 and CP2 as two *necessary* and *sufficient* conditions for the *correctness* of transformation functions or even an OT solution as a whole, which had misled people to treat CP1 and CP2 as two golden rules for evaluating the correctness of OT. In fact, CP1 and CP2 are *neither* necessary *nor* sufficient for the correctness of transformation functions, let alone for the correctness of a whole OT solution. CP1 and CP2 are *unnecessary* for transformation functions because they can be avoided by using OT control algorithms (e.g. CP2-avoidance algorithms [10,24,32,34,45,46,63,71,72,78]). CP1 and CP2 are *insufficient* because they govern only *convergence*, but not *intention preservation* (e.g. the *combined effects* of concurrent operations in text editing [66,67]) in co-editors [51]. Without intention preservation, transformation functions can preserve CP1 and CP2 *trivially*. For example, the function *trivial-TF(O, $O_x$) = O′* transforms O against $O_x$ to produce O′, where O′ is an operation that always replaces existing contents of the document with a number *X*. It can be shown that this *trivial-TF* preserves CP1 and CP2. By assigning *X* to an arbitrary number, one can get an infinite number of transformation functions capable of preserving CP1 and CP2, but none of them is *meaningful*, let alone be *correct* for co-editing.

Another symptom is to attribute the root cause of every OT puzzle that may result in document divergence to CP2-violation. In fact, divergence is only a *symptom* of a puzzle and could be *caused* by violating transformation conditions other than CP2 [57]. For example, the dOPT puzzle could result in divergent states, but it was actually caused by violation of the *context-equivalence* condition [50], rather than CP2[13]. However, after the dOPT puzzle had been resolved for over a decade, we still see publications (e.g. [39,42]) attributing this puzzle to *CP2-violation* and trying to relate CP2 solutions (e.g. TTF [39]) to the dOPT puzzle. Accurate attribution of an OT puzzle to the root transformation condition is crucial not only to resolving the specific puzzle, but also to designing and evaluating the correctness of OT solutions in general. The incorrect attribution of the dOPT puzzle to CP2-violation reflects a misunderstanding of fundamental OT correctness conditions (e.g. context-based conditions [50,63,78]).

---
[12] In [40], CP1 and CP2 are named as TP1 (Transformation Property 1) and TP2 (Transformation Property 1), respectively.
[13] The essential reason for the adOPTed algorithm to resolve the dOPT puzzle is its capability of ensuring the *context-equivalence* condition [50], rather than requiring transformation functions to preserve CP1 and CP2 (though the two properties are important for different purposes). The *context-equivalence* condition was implied in the description of the adOPTed algorithm but not explicitly stated in [40], whereas the CP1-CP2-theorem was explicit, which was often misinterpreted as the reason for the adOPTed algorithm to resolve the dOPT puzzle.

Yet another symptom is to label OT solutions as being *incorrect* on the ground of *not* preserving CP2 [4,5,7,19,37,39,42,75]. Arguments along this line could be traced back to the early history of exploring alternative solutions to the FT puzzle (a case of CP2-violation). After the discovery of the FT puzzle in [49,51], numerous attempts were made to resolve this puzzle, resulting in a large number of proposals [19,20,25,26,27,28,37,38,39,48]. Though different to each other, those proposals share one common characteristic: they made radical changes to the core components of OT, e.g. control algorithms or data/operation models, which are fundamental to an OT solution, but have little to do with the FT puzzle. Consequently, such changes often brought in new correctness and efficiency issues that were more complex than the original FT problem they were proposed to solve. There were numerous publications claiming to have *disproved* all prior (by then) OT solutions in terms of CP2-correctness (e.g. using theorem provers), and proposed new solutions that were *verified* to be CP2-correct (using the same theorem provers); but those proposals or verifications were repeatedly found to be flawed later (e.g. see counter-examples in [19,25,27,28,48]). Erroneous results from those attempts had the effect of creating the illusion that OT was full of puzzles that were spiraling out of control, which had caused major confusions among practitioners and later researchers entering the field.

To help clear up these misconceptions, we highlight the following facts. First, despite a variety of CP2-violation puzzles reported in literature, all reported CP2 puzzles were just *variations* of the same basic FT puzzle or *derivatives* of erroneous solutions proposed to solve the original FT puzzle [57]. Second, nearly all the *FT-solution* attempts had been confined to a primitive operation model for text editing with *character*-wise *insert* and *delete* operations. Based on exhaustive examination of all possible transformation cases under this primitive model, it has been proven that the FT puzzle is the *only* possible CP2-violation case in OT solutions supporting commonly adopted combined-effects for pair-wise concurrent operations in text editing [66]. Last and the most important fact is that all possible CP2-violation cases under a more general *string*-wise operation model[14] for text editing have been detected and solved by verified (and efficient) solutions based on *CP2-preservation* (applicable to text editing) and *CP2-avoidance* (applicable beyond text editing) strategies [24,34,45,63,66,67,71,72,78], as discussed in Section 4.1.3.

*4.3.2. Twin Solutions to CP2-Violation: WOOT and TTF*

Among various alternative proposals (other than the ones summarized in Section 4.1.3) to address the CP2-violation issue, a pair of solutions are particularly noteworthy: one is the WOOT solution, and the other is the TTF (*Tombstone Transformation Functions*) solution, both of which were based on the same notion of *tombstone-based* object sequences, and proposed at nearly the same time by the same authors [37,38,39]. WOOT was proposed as an OT-alternative capable of avoiding the CP2-violation issue, and became the *first* of CRDT solutions for text co-editing [4,5,7,23,30,36,37,38,42,43,44,73,74, 75]; and the TTF solution was claimed to be the *first* and often cited as the *sole* OT solution capable of preserving CP2 [4,5,7,19,37,39,42, 75]. As such, the TTF solution was often used as the representative OT solution in comparison with CRDT solutions. Quite some claims about CRDT superiority over OT were based on the comparison between CRDT solutions (e.g. Logoot, RGA, WOOT variations, etc.) and the TTF solution (typically integrated with the SOCT2 control algorithm [47]). For example, the TTF solution was reported to be outperformed by Logoot for a factor up to 1000 in [4]. This *1000-times-gain* claim was widely cited as an experimental evidence (e.g. footnote 8) for CRDT's performance superiority over OT [4,5,7,42,75]. While validating the Logoot's *1000-time-gain* over TTF is outside the scope of this paper, what we want to point out here is that those CRDT and TTF claims are *groundless* because: (1) they are *contrary* to the *facts* that numerous OT solutions have been proven to be correct with respect to well-established conditions and properties (including CP1 and CP2) *before* and *after* TTF and WOOT solutions [10,24,34,45,63,66,67,71,72,78]; and (2) they are also *mistaken* about what the TTF solution *really* is − a *hybrid* of CRDT and OT, as elaborated below.

---

[14] In [67], an additional False-Border (FB) puzzle under a pair of string-wise insert and delete operations was detected and resolved, and it was shown that the FT and FB puzzles were the only two possible CP2-violation cases in OT solutions supporting string-wise operations (with commonly adopted combined effects of concurrent operations) in text co-editors.

In the TTF solution, an internal *tombstone-based* object sequence is maintained, which is a characteristic CRDT component (like WOOT). In addition, an existing OT control algorithm (e.g. SOCT2 [47]) and special CP2-preserving transformation functions (i.e. TTF [39], defined for a pair of character-wise *insert* and *delete* operations on a sequence of objects with tombstones) were used to transform operations, which is similar to OT. One subtle but crucial detail deserves attention: operations being transformed by TTF functions are *not* user-generated operations as in typical OT solutions, but *internal operations* which are defined on and only applicable to the *internal object sequence*. Consequently, additional conversions between internal and external operations are required, which is typical to CRDT solutions (like WOOT). Due to its hybrid nature, the TTF solution bears the costs of both CRDT and OT, with the main costs dominated by its CRDT components, including the maintenance of the tombstone-based object sequence and associated schemes (each with the time complexity $O(C_t)$) for converting between internal and external operations. We refrain from detailed comparison of TTF with OT or CRDT in this paper, but will present comprehensive comparisons of OT, CRDT, TTF, and other alternatives (including those in [17,25,26,27,28]), that are based on the same general transformation approach in future papers.

*4.3.3. Differences between OT and CRDT in Time and Space Complexity*

For comparison convenience, we have summarized the time and space complexities of representative OT and CRDT solutions (with references to those solutions) in Table 4. The results in this table disprove CRDT superiority claims over OT in time and space complexity.

Table 4 Space and time complexities of representative (not exhaustive) OT and CRDT solutions. $m$ (usually $1 < m \leq 5$) is the number of users in a real-time co-editing session.

|  | **OT**<br>[24,32,34,40,45,46,47,50,51,63,71,72,78] | **CRDT** | |
|---|---|---|---|
|  |  | **Tombstone-based**<br>**WOOT** variations [4,38,74] +<br>**RGA** [42] | **Non-tombstone-based**<br>**Logoot** variations [5,73,75] |
| **Space** | $O(c)$ [24,32,34,45,46,71,72],<br>$O(c*m)$ [47,49,50,51,78], or<br>$O(c*m^2)$ [40,63] | $O(C_t)$ for WOOT variations<br>$O(C)$ for RGA | $O(C)$ to $O(C^2)$ for all Logoot variations |
| **Time** | Local: $O(1)$ for all referred OT solutions,<br>Remote: $O(c)$[24,32,34,45,46,63,71,72,78]<br>or $O(c^2)$ [40,47,50,51] | Local & remote:<br>$O(C)$ for RGA,<br>$O(C_t^2)$ [4,74], or<br>$O(C_t^3)$ [38] | For all Logoot variations:<br>Local: $O(C)$<br>Remote: $O(C \circ log(C))$ |

Furthermore, the *real* complexity differences between OT and CRDT solutions should be examined not only by *theoretic* differences using the big-*O* notation, but also by *practical* differences of the input variables in those expressions: *c* is often bounded by a small value, e.g. $0 \leq c \leq 10$, for real-time sessions with a few users; *C* is orders of magnitude larger than *c*, e.g. $10^3 \leq C \leq 10^6$, for common text document sizes ranging from 1K to 1M characters, while $C_t$ is much larger than $C$ ($C_t \gg C$) with the inclusion of tombstones. These practical differences are often more significant than the theoretic differences to real world co-editing applications.

Finally, two characteristic differences between OT and CRDT in time and space costs are noteworthy. First, OT has no cost for transformation time and the operation buffer is empty (with garbage collection) when there is no concurrent operation, whereas CRDT bears the same space and time costs regardless whether operations are *sequential* or *concurrent* as their effects always have to be recorded and kept in the object sequence, which means the same cost has to be paid even if there is no concurrent editing. Second, OT has no transformation cost in handling a local operation since a local operation can never be concurrent with any operation in the buffer, whereas CRDT has almost the same processing costs regardless whether an operation is *local* or *remote,* which could have adversary impact on the local responsiveness of CRDT-based co-editors [7].

*4.3.4. Simplicity of CRDT vs OT*

In terms of time and space complexity, $C_t/C$–based CRDT solutions are clearly more complex than *c*-based OT solutions, as summarized in Table 4. However, one often-cited CRDT merit is its

*simplicity*. This section tackles various versions and arguments on simplicity of CRDT, which we have found scattered in published literature [4,7,36,43,44,75] and discussions among developers.

One version of the CRDT simplicity argument can be sketched as follows: CRDT works without OT, thus avoids complex issues with OT, hence CRDT is simple. The fallacies of this argument are: avoiding the issues of an existing approach may *not* make the new approach simple; and the simplicity of one approach is *not* determined by whether it works without the other approach (obviously, OT works without CRDT as well). The relevant questions that should be asked are: *what* special issues each approach has, *whether* those issues are easy to solve and have been solved, and *whether* solutions to such issues are simple. In previous sections, we have provided ample evidences that CRDT has its own challenging issues and many of them remain unsolved (see elaborations in Section 4.2); and CRDT solutions are *not* simple but more *complex* than OT solutions, as shown in Table 4.

Another simplicity argument goes as follows: CRDT makes concurrent operations *natively* commutative, whereas OT makes concurrent operations commutative *after the fact*, so the CRDT approach is more *elegant* and *simpler* than OT [43,44]. Unfortunately, this simplicity augment is misleading because: CRDT identifier-based operations are *not* native to editors; CRDT is *not* different from OT in making non-commutative position-based operations commutative in editors after the fact (albeit *indirectly*); and the CRDT approach to making position-based operations commutative is more complex than OT (see Section 4.2 and Table 4).

Yet another CRDT simplicity has been argued in relation to implementation. Based on our experiences of designing and implementing over a dozen of OT-based co-editing prototypes and production systems, in desktop, Web, and mobile platforms, we have learnt that the bulk of the implementation challenges lie in *how* to apply OT solutions in the context of a real-world editing system (see Section 5), rather than implementing OT algorithms themselves, which are at the core *but* only part of a co-editing system. OT was reported to be hard to implement in a well-known quote by a former Google Wave engineer[15]:

> *"Unfortunately, implementing OT sucks. [...] The algorithms are really hard and time consuming to implement correctly. [...] Wave took 2 years to write and if we rewrote it today, it would take almost as long to write a second time."*

The above quote was widely cited, and used by some people to dismiss OT, and to argue for CRDT simplicity by following the logic − *what is hard for OT will be easy for CRDT* (as CRDT works without OT). However, what is less known and ignored is that the same engineer later amended the previous comments with[16]:

> *"For what its worth, I no longer believe that wave would take 2 years to implement now - mostly because of advances in web frameworks and web browsers. When wave was written we didn't have websockets, IE9 was quite a new browser. We've come a really long way in the last few years."*

The above amended statements revealed that major challenges of Google Wave were due to the Web frameworks, browsers, and communication utilities used to build the whole OT-based Google Wave, rather than just implementing the core OT algorithms [34,72]. This reflection is consistent with our experiences. The idea that CRDT is simple to implement is unfortunately not substantiated by evidence, but contradicted by the fact that CRDT implementations in working co-editors are rarely seen and robust implementations are virtually nonexistent (see the next section).

The basic ideas and external effects of OT are simple to illustrate (see Fig. 1-(a) and (b)), but inner-workings of OT are not simple to understand by non-experts or application developers. There is still large room for improvement in making OT solutions more accessible to practitioners [46,57,59,70], and most importantly in applying OT to real world collaborative applications, which is the subject of the next section.

---

[15] https://www.championtutor.com:7004/ (Nov 6, 2011)

[16] https://news.ycombinator.com/item?id=12311984 (Aug 18, 2016)

# 5 BUILDING CO-EDITORS BASED ON OT AND CRDT

Building and experimenting with working co-editors has played a crucial role in advancing co-editing research for over two decades. The methodology and practice of experimentation through *working* co-editors has helped the research community to uncover intricate technical issues that would otherwise go unnoticed by pure theoretical study, to experimentally validate proposed solutions, and to gain critical insights for deriving general principles and theories, which in turn inspire and guide experimental exploration. The success of co-editing research in adoption by industry products owes in no small part to the research community's relentless efforts in connecting theory with practice and in innovating system design and implementation solutions. In this section, we examine the role of system implementation played in OT and CRDT research, and the consequential differences in OT and CRDT adoption in real world co-editors and industry products.

## *5.1  OT-based Co-Editors*

One primary goal of OT research is to invent innovative solutions that can be used for building useful and useable co-editors. While theoretic work around OT *algorithms* has been fundamental to this endeavor, the practice of building OT-based working co-editors has historically played a crucial and integral role in driving and shaping OT solutions. The very first OT research publication detailed the design and implementation of a plain-text co-editor GROVE [12]. A succession of working co-editors, including DistEdit [35], Jupiter [34], JOINT EMACS [40], and REDUCE [49,51,54], were built by researchers to investigate both system and theoretical issues in co-editing. These experimental efforts revealed critical insights into the dOPT puzzle, and eventually led to resolving it and establishing the theoretic foundation for OT − a comprehensive set of transformation conditions and properties, such as context-based conditions and CP1 and CP2 properties [35,40,50,51,57].

Early OT-based co-editors served as research vehicles to investigate novel concurrency control techniques for co-editing plain-text documents, but placed little emphasis on the relevance to supporting real world applications that users may use daily for content creation [16]. It was around the year 2000 that researchers began to investigate the possibility of extending OT from supporting plain-text documents to off-the-shelf productivity suites with complex document formats and comprehensive functionalities. One representative work along this line of investigation is the *Transparent Adaptation* (TA) approach [56,61,76]. The goal of the TA approach was to extend the basic OT to support complex applications and to convert single-user editors into co-editors, without changing the source code of the original applications. A set of diverse productivity applications were examined and successfully converted into co-editors, including Microsoft Word (CoWord [56,61,64,76]), PowerPoint (CoPowerPoint [56,61]), and Autodesk Maya (CoMaya [1,2]).

A cornerstone of TA is the *concurrency-centric* nature of OT, which keeps the core OT control algorithms generic and allows transformation capabilities to be extensible to new application domains. The TA approach is able to handle a myriad of user interactions and complex data objects found in modern productivity applications for real-time co-editing. More specifically, a TA-based co-editor consists of three architectural components: (1) Generic Collaboration Engine (GCE), which provides generic transformation capabilities (independent of specific OT algorithms or transformation function); (2) Collaboration Adaptor (CA), which bridges the GCE with an existing single-user application, effectively extending basic transformation functionalities to the target application; (3) A single-user editor, which provides conventional editing interface features and functionalities to users and suitable API (Application Programming Interface) to the CA. With the TA approach and a reusable GCE, the task of building a new co-editor is reduced to building a new CA for supporting the special data and operation models of an existing single-user editor, without reinventing a new editor or re-implementing an OT-based collaboration engine.

Working co-editors such as CoWord, CoPowerPoint and CoMaya demonstrated OT relevance to real world applications and helped bridge the research community and industry [31]. The research community and members of industry have since jointly organized a series of co-editing workshops [33], to share experiences in building and using co-editing systems. Researchers have given technical talks at industrial labs [55,58], demonstrated working prototypes [55,58,60], and delivered tutorials on co-editing technologies [59] at ACM CSCW conferences.

### 5.2 OT-based Co-Editing Products

In 2009, Google announced adoption of OT as the core technology for supporting its real-time collaboration features in Google Wave [17] [72,70]. Google Wave OT algorithms and protocols were handed over to the Apache Software Foundation and open sourced under the name Apache Wave. It has had strong influences on a number of Web-based open-source OT software projects. One representative open source OT project is ShareJS[18], which was led by a former Google Wave engineer. In 2010, OT was adopted in Google Docs [10] – a web-based real-time collaborative office suite. Other notable OT-based co-editors include SubEthaEdit[19], CKEditor[20], Etherpad[21], to name a few.

In recent years, cloud storage companies started to extend their storage and file synchronization services to offering new Web-based co-editing services on the files in their storage. These companies built their own OT-based rich-text co-editors, such as Dropbox Paper[22] (2017), and Box Notes[23] (2017). More recently (2018), Tencent also integrated the OT-based rich-text co-editing capability into its cloud-based TAPD (Tencent Agile Product Development) environment[24].

In contrast to Google Docs and other Web-based co-editors that were built from scratch, Codox supports real-time co-editing directly in existing Web-based applications, such as Gmail [25], Evernote (see footnote 25), WordPress[26], Zendesk[27], Wikipedia[28], TinyMCE[29], and Quill[30], and retain functionalities and the "look-and-feel" of original Web applications. In each of Codox co-editing applications, an application-specific adaptor is injected into the single-user editing environment, which bridges an existing editor and a generic OT-powered Codox engine. The Codox approach has been directly inspired by academic research on OT and TA [1,2,9,50,51,52,53,54,56,57,61,62,63,64,65,66,67,76,77,78].

Last decade has witnessed significant efforts from industry and open source communities in applying OT to numerous real world applications, with large scale deployment. The theory and practice of OT is now driven by a vibrant community of academic researchers and industrial practitioners. A wealth of valuable experiences have been accumulated from building and using co-editing applications. This calls for a comprehensive technical review of real world system approaches, issues and experiences, which we plan to do in the future.

### 5.3 CRDT-based Co-Editors

Began as an effort to address the FT puzzle in OT, CRDT research for co-editing has adopted predominantly theoretic approaches to studying co-editing issues and verifying proposed solutions (e.g. using theorem provers, or model checkers) [19,20,37,38,39], but rarely implemented and validated CRDT solutions in working co-editors. These approaches have had profound impact in shaping CRDT research and solutions. We found no working CRDT-based co-editors built by CRDT researchers or academic literature documenting *system* experiences in using proposed CRDT solutions for building working co-editors.

---

[17] https://en.wikipedia.org/wiki/Apache_Wave
[18] https://github.com/josephg/ShareJS
[19] https://www.codingmonkeys.de/subethaedit/
[20] https://ckeditor.com/collaborative-editing/
[21] http://etherpad.org/
[22] https://paper.dropbox.com/
[23] https://www.box.com/notes
[24] https://www.tapd.cn/
[25] https://chrome.google.com/webstore/detail/wave-for-gmail-and-everno/dggkchdpgkalbmhnlmgmiafjacofjghb?utm_source=inline-install-disabled
[26] https://wordpress.org/plugins/wave-for-wp/
[27] https://www.zendesk.com/apps/support/wave/
[28] https://www.wikipedia.org/
[29] https://codepen.io/dnus/pen/ELRNMo
[30] https://codepen.io/dnus/pen/OojaeN

We did however encounter a dozen of CRDT-based plain-text co-editor projects hosted on GitHub[31], which were created by non-academic practitioners who were interested in learning whether and how CRDT solutions actually worked when applied to realistic editing environments. Most of those implementations are at rudimentary stages of development, but we found two relative stable prototypes: Teletype[32], which is based on WOOT (with tombstones) [7,37,38], and Alchemy Book (see footnote 10), which is based on Logoot (without tombstones) [73,75]. Based on experimentation with live demos and review of available documentation and source code of those prototypes hosted at GitHub, we can make a number of observations.

First, *CRDT-based co-editors were mostly developed by combining a CRDT solution with an existing text editor*. For example, Teletype was developed by integrating WOOT with a desktop text editor named Atom[33]; Alchemy Book was built by integrating Logoot with a Web-based text editor named CodeMirror[34]. The implementations of Teletype and Alchemy Book invoke similarities with the TA approach [56]. These similarities are unsurprising since TA is based on the general transformation approach but independent of specific transformation solutions, and CRDT is shown to be an instance of the general transformation approach.

Second, *concrete implementations of CRDT in co-editors revealed key missing steps in CRDT literature*. We carefully examined how co-editing is supported *end-to-end* under CRDT-based co-editors, i.e. from the point when an operation is generated from a local document by a user, all the way to the point when this operation is applied to a remote document seen by another user. Both Teletype and Alchemy Book, as well as other CRDT-based co-editor implementations, unmistakably convert user-generated *position-based* operations into *identifier-based* operations at local sites (this is *obscured* in Logoot), and convert remote *identifier-based* operations, after applying in internal object sequences, to *position-based* operations at remote sites (this is *ignored* in WOOT and RGA). This observation confirms our illustration in Fig. 1-(c) and description of CRDT under the general transformation framework in Section 3. It should be stressed that these missing steps are *not* mere implementation details, but *necessary* and *crucial* steps for CRDT solutions to achieve consistency maintenance in co-editors.

Third, *all working CRDT-based co-editing systems have used a central server to support some aspects of co-editing*. For example, Alchemy Book uses a central server for session management and broadcasting messages among co-editing clients, which is similar to OT-based CoWord [56]; Teletype uses a central server for client-discovery (session management) but allows co-editing clients to be *fully connected* for broadcasting messages without involving the server, which is similar to OT-based REDUCE [51,54]. To our best knowledge, there has been no single example of CRDT-based co-editors that were built without using a client-server architecture. The often-suggested idea that CRDT solutions are specifically designed for peer-to-peer collaborative editing [5,38,42,73,74,75] is tenuous at best, and confounds what CRDT solutions like WOOT and Logoot actually do. We further elaborate this point in Section 5.5.

By experimenting with Teletype and Alchemy Book prototypes, we can confirm the analytical results about *tombstone-based* (WOOT) and *non-tombstone-based* (Logoot) CRDT variations in Section 4.2. In Teletype, we experienced WOOT-based *tombstone overhead* effects, where the co-editor suffered significant increase of memory and degradation of performance, in both local response and remote replay, as the number of deletions increases during a session. In Alchemy Book, we were able to produce *concurrent-insert-random-interleaving* results when performing concurrent insertions at the same location, and experienced document inconsistencies under numerous scenarios (Fig 2). The insert interleaving abnormality was also independently reported by a developer who tried to implement Logoot for text co-editing[35].

---

[32] https://github.com/atom/teletype
[33] https://atom.io/
[34] https://codemirror.net/
[35] https://stackoverflow.com/questions/45722742/logoot-crdt-interleaving-of-data-on-concurrent-edits-to-the-same-spot

## 5.4  CRDT-based Co-Editing Products

Among CRDT-based working co-editors, Teletype for Atom is sometimes cited in the CRDT community as an example of industrial adoption of CRDT in co-editors. Apart from Teletype, we are not aware of any other industry co-editing product that is based on a CRDT solution for consistency maintenance.

Why CRDT solutions were rarely adopted in industrial co-editing products? Apart from the CRDT correctness and efficiency issues discussed before, in our view, another main obstacle to CRDT adoption is that CRDT solutions for co-editors are mostly confined to inserting and deleting characters or objects in a linear sequence, which are clearly inadequate for supporting real world applications, e.g. rich text formatting, tables, images, etc., which have become a de facto requirement for collaborative content creation and supported by co-editing products, such as Google Docs, Dropbox Paper and Codox Apps, etc. Moreover, key functional components of existing CRDT co-editing solutions are *intricately* coupled with specific object sequences and operations, which are not re-usable and have to be re-invented for every application with data and operation models different from text editing.

## 5.5  Myths and Facts about Peer-to-Peer Co-Editing

As pointed out before, all known co-editors have used the client-server model for supporting some aspects of co-editing, e.g. discovering and tracking users in a session, and/or operation broadcasting. In spite of this, academic work on CRDT has been using "peer-to-peer collaborative editing" ("p2p co-editing" in short) as one primary motivator and differentiator between CRDT and OT [5,38,42,73,74,75]. We have found that the term of "p2p co-editing" in existing literature is ill-defined and often conflates multiple factors. In this section, we will tease apart these factors and discuss their relationships to OT and CRDT solutions.

The first factor is *what constraints or conditions are imposed (by OT or CRDT) on operation propagation and communication in co-editors*. Generally, both OT and CRDT solutions require editing operations to be executed in their *causal-effect* orders (based on the *happen-before* relation [22]) at all co-editing sites, to meet the *causality-preservation* requirement for co-editors [12,50,51]. This causal ordering can be achieved by adopting any suitable distributed computing techniques (based on either client-server or peer-to-peer models), which are *orthogonal* to OT or CRDT. However, some CRDT solutions, like WOOT, replaced the general causal ordering condition with WOOT-specific execution conditions: an operation can be accepted for execution only if the two neighboring objects (for an insert) or the target object (for a delete) already exist in the internal object sequence. WOOT-specific execution conditions were quoted as a merit for supporting p2p co-editing [38], but such conditions cannot ensure *causality-preservation* generally required for co-editors [12,50,51], and they are costly to check with the time complexity $O(C_t)$. Other CRDT solutions (e.g. RGA and Logoot variations) and OT solutions are the same in adopting the general causal ordering condition. Hence, with the exception of WOOT variations, causally-ordered operation propagation and execution are common requirements for OT and CRDT solutions, and hence are not, in general, differentiating factors between OT and CRDT.

The second factor is *whether any server is required (by OT or CRDT) in operation propagation and communication*. CRDT solutions do not explicitly require the existence of a communication server, but with the exception of WOOT, assume by default the existence of an external *causal-order-preserving* communication *service*. On the other hand, some OT solutions, notably Jupiter [34], NICE [45], and Google Wave and Docs [10, 72], explicitly require a central *transformation-based* server to do part of the transformation and to broadcast operations; SOCT3 and SOCT4 [71] require a special server to issue continuous total ordering numbers for labelling operations. However, most other OT solutions, such as adOPTed [40], GOT[49,51], GOTO[50], SOCT2 [47], TIBOT [24,78], COT [62,63], and POT[78], do not require a central server to do (any part of) the OT work, but only require the use of an external *causal-order-preserving* communication *service* (the same as most CRDT solutions). Therefore, whether or not requiring a server is also not a general differentiating factor between OT and CRDT either.

Yet another factor often cited in connection to p2p co-editing is whether vectors (with one element for each of co-editing sites in a session) or scalars (with a fixed number of variables) are used for timestamping or control in OT or CRDT solutions. Again, both vector-based and scalar-based timestamps and control have been used in both OT and CRDT solutions. For example, the RGA solution uses vector-based timestamps to reduce the time complexity from $O(C_t^2)$ (in WOOTH and WOOTO) to $O(C_t)$ and for garbage collection of tombstones (further reduce $O(C_t)$ to $O(C)$); Logoot variations use variable length (bounded by the object sequence length $C$) object identifiers, and require an external causally-ordered broadcasting service. On the other hand, some OT solutions, including adOPTed [40], GOT [49, 51], GOTO [50], and COT [62,63], used vector-based timestamps, but other OT solutions, including Jupiter [34], NICE [45], TIBOT [24,78], and SOCT4 [71], Google Wave and Docs [10,72], and POT [78], used scalar-based timestamps. In fact, scalar-based timestamping had been introduced to OT solutions long before the first CRDT solution (WOOT) appeared. Therefore, neither CRDT nor OT is unique in using vector-based and scalar-based timestamps or control.

In summary, *all* p2p co-editing factors are *orthogonal* to OT and CRDT and cannot differentiate OT and CRDT in general. All co-editing systems built so far have been based on client-server architectures, and it remains open whether, what and how to support p2p co-editing in the future.

## 6 CONCLUSIONS

In this work, we have conducted a comprehensive review and comparison of OT and CRDT for consistency maintenance in real-time co-editing and in building real world co-editors. We have made a number of discoveries, which contribute to the advancement of the state-of-the-art knowledge on collaboration-enabling technology in general, and on OT and CRDT in particular.

One significant outcome of this study is the discovery of the general transformation approach, which not only provides a common ground for describing, examining and comparing a variety of concurrency control solutions in co-editing (e.g. OT and CRDT solutions, among others), and also may inspire invention of new concurrency control solutions in co-editing in the future.

Another significant outcome is that: CRDT is the *same* as OT in following the general transformation approach to real-time co-editing; CRDT is the *same* as OT in making user-generated operations commutative *after the fact*, albeit *indirectly* (by CRDT) rather than *directly* (by OT); and CRDT operations are *not* native or commutative to text editors, but require additional conversions between CRDT internal operations and external editing operations. Revealing these previously hidden by critical facts helps demystify what CRDT really *is* and *isn't* to co-editing, and in turn bring out the *real differences* between OT and CRDT for co-editors − their radically different ways of realizing the same general transformation approach.

One key insight from probing what really differentiates OT and CRDT is: OT is *concurrency-centric* in the sense it treats *generic* concurrency issues among operations as its *first priority* at the core control algorithms, and *isolates* the handling of *application-specific* data and operation modelling issues in transformation functions; whereas CRDT is *content-centric* in the sense that it takes the *first priority* to manipulate internal application related *contents*, including object sequences and schemes for searching and applying identifier-based operations in the object sequence, but *mixes* the handling of concurrency issues within object search and manipulation schemes. This *concurrency-centric* vs *content-centric* difference is fundamental and has profound implications to OT and CRDT solutions.

The first significant implication is found in the different design issues and challenges in OT and CRDT solutions. Key OT design issues include designing control algorithms to deal with *generic* concurrency issues, and designing *separate* transformation functions to handle *application-specific* issues; OT-special challenges and puzzles (all solved), such as ensuring context-based conditions (e.g. the *dOPT* puzzle was a case of violating the *context-equivalence* condition), and avoiding or preserving CP2 (e.g. the *FT* puzzle was a case of violating of the *CP2 property*), were derived from and solved under the concurrency-centric approach. In contrast, key CRDT design issues include designing CRDT-special data structures and schemes for representing and manipulating object sequences, searching and executing identifier-based operations in the object sequence, and conversions between internal *identifier-based* operations and external *position-based* operations, which *collectively* deal with both application-specific and concurrency issues in co-editing. This approach has induced a myriad of CRDT-specific challenges and puzzles, such as *the correctness of key CRDT data structures and functional components*, *tombstone overhead*, *variable and lengthy identifiers*, *inconsistent-position-integer-ordering and infinite loop flaws*, *position-order-violation puzzles*, and *concurrent-insert-interleaving puzzles*. It remains an open challenge to resolve these issues under the CRDT approach to co-editing.

The second significant implication is found in the different time and space *complexities* among OT and CRDT solutions. OT complexity is determined by a variable $c$ (for *concurrency*) – the number of concurrent operations involved in transforming an operation; CRDT complexity is dominated by a variable $C$ (for *Contents*) or $C_t$ (for *Content with tombstones*) – the number of objects maintained in the internal object sequence. In terms of theoretic complexity (see details in Table 4), representative OT solutions have achieved the time complexity $O(1)$, and $O(c)$ or $O(c^2)$ for processing local and remote operations, respectively; and the space complexity $O(c)$, $O(c*m)$, or $O(c*m^2)$, where $m$ is the number of real-time co-editing users in a session (usually < 5). In contrast, representative CRDT solutions have the time complexity ranging from $O(C_t^3)$, $O(C_t^2)$, to $O(C)$ (for tombstone-based solutions), or $O(C \circ log(C))$ (for non-tombstone-based solutions); and the space complexity ranging from $O(C_t)$ or $O(C)$ (for tombstone-based solutions, with or without tombstone garbage collection), to between $O(C)$ and $O(C^2)$ (for non-tombstone-based solutions).

In addition to examining the *theoretic* complexity differences, we highlight the *practical* differences of the input variables in those complexity expressions: $c$ is often bounded by a small value, e.g. $0 \leq c \leq 10$, for a real-time session with a few users; $C$ is orders of magnitude larger than $c$, e.g. $10^3 \leq C \leq 10^6$, for common plain text document sizes ranging from 1K to 1M characters, while $C_t$ could be much larger than $C$ with the inclusion of tombstones. In real-time text co-editing, the following *inequality* commonly holds: $C_t \gg C \gg c$. It remains an open challenge to devise $C_t/C$-based CRDT solutions that are superior over $c$-based OT solutions in time and space complexity and in practical performance.

The third implication is in the *generality* and *extendibility* of OT and CRDT solutions for co-editors. OT solutions separate generic concurrency issues from application-specific data and operation issues, with the core control algorithms being generally applicable to different application domains beyond text editing. Extensions of existing OT solutions can be and have been achieved by designing new transformation functions for new applications, without reinventing its core control algorithms. In contrast, CRDT solutions mix concurrency issues with application-specific data and operation issues, with key CRDT components being intricately related to each other and coupled with application-specific object sequences and operations. So far, CRDT for co-editing has been confined to plain-text editing; it is unknown whether CRDT solutions can be extended for supporting real world co-editors beyond plain-text editing.

In addition to the above significant differences between OT and CRDT, one more major difference lies in *implementation* and *validation*: numerous OT implementations have been validated in working co-editors (and made available in open source or commercial distributions), whereas CRDT solutions were rarely implemented and validated in working co-editors. We believe these differences are the key factors that have affected the adoption of and choice between OT and CRDT for co-editors in the real world.

In this work, we have critically reviewed and evaluated representative OT and CRDT solutions, with respect to *correctness*, *time and space complexity*, *simplicity*, *applicability in real world co-editors*, and *suitability in peer-to-peer co-editing*. The evidences and evaluation results from this work disprove superiority claims made by CRDT over OT on all accounts.

In over two decades, co-editing research has evolved from a niche area of in CSCW and distributed computing to a host of core collaboration-enabling techniques widely used in real world co-editors and major industrial co-editing products. Along the way, co-editing research has resolved numerous theoretic, technical, and application challenges, and established core co-editing techniques (e.g. for consistency maintenance, among others) with theoretically verified and experimentally validated correctness and high efficiency. Moving forward, we foresee open challenges and new opportunities in bridging core co-editing techniques with emerging and co-editing applications, and in extending core techniques to application domains beyond co-editing.

We hope discoveries from this work will help clear up myths and common misconceptions surrounding alternative co-editing approaches and techniques, inspire new explorations of novel collaboration techniques, and accelerate progress in co-editing and collaboration-enabling technology innovation and real world applications.

# 7 ACKNOWLEGMENTS

This research is partially supported by an Academic Research Fund Tier 2 Grant (MOE2015-T2-1-087) from Ministry of Education Singapore.